# Room temperature Multiferroicity and Magnetoelectric coupling in Ca/Mn modified BaTiO$_3$

P. Maneesha[1], Koyal Suman Samantaray[1], Rakhi Saha[1], Tabinda Nabi[1], Rajashri Urkude[2], Biplab Ghosh[2], Arjun K Pathak[3], Indranil Bhaumik[4,5], Abdelkrim Mekki[6,7], Khalil Harrabi[6,8], Somaditya Sen[1]*

[1]*Department of Physics, Indian Institute of Technology Indore, Indore, 453552, India*
[2]*Beamline Development & Application Section, Bhabha Atomic Research Centre, Trombay, Mumbai 400085*
[3]*State University at New York, Buffalo State, NY, USA*
[4]*Crystal Growth Laboratory, High Energy Lasers and Optics Division, RRCAT, Indore- 452013, India*
[5] *Homi Bhabha National Institute, Training School Complex, Anushakti Nagar, Mumbai-400094, India*
[6] *Department of Physics, King Fahd University of Petroleum & Minerals Dhahran, 31261, Saudi Arabia*
[7]*Interdiciplinary Research Center (IRC) for Advanced Material, King Fahd University of Petroleum & Minerals, Dhahran 31261, Saudi Arabia*
[8]*Interdisciplinary Research Center (RC) for Intelligent Secure Systems, King Fahd University of Petroleum & Minerals, Dhahran 31261, Saudi Arabia*

*Corresponding author: sens@iiti.ac.in

## Abstract

Materials with magnetoelectric coupling (MEC) between ferroic orders at room temperature are emerging field in modern technology and physics. BaTiO$_3$ is a robust ferroelectric in which several doping has led to MEC. In Ca and Mn modified BaTiO$_3$ has been study with a series of Ba$_{(1-x)}$Ca$_{(x)}$Ti$_{(1-y)}$Mn$_{(y)}$O$_3$ (x=y= 0, 0.03, 0.06, 0.09), in this MEC was only observed in x=0.03. The structural modifications with changing substitution reveal a reduced Ti-O-Ti bond angle for this sample which is the most ferromagnetic in nature. A mixed phase of tetragonal *P4mm* and hexagonal *P6$_3$/mmc* space groups of BaTiO$_3$ is observed in the substituted samples, with nominal contribution of the hexagonal phase for x=0.03. A valence state study using XPS and XANES reveals the presence of enhanced proportion of Mn$^{3+}$ ions in the sample which support a pseudo Jahn-Teller distortion, thereby supporting the ferroelectricity for x=0.03. Direct evidences of MEC was obtained from magnetoelectric measurements. A magnetoelectric coupling coefficient, $\alpha_{ME}$ ~44 mVcm$^{-1}$Oe$^{-1}$ was obtained for dc magnetic field of 600 Oe and a 10Hz ac field of 40 Oe. Such MEC was not observed for higher substitution which emphasizes the sensitivity of the structural properties on substitution.

## Introduction

Magnetoelectric (ME), multiferroic materials are materials required for next generation novel multifunctional devices. Hence, these are high priority research materials in the ferroics community. Experimentally discovered about sixty years ago, ME effect has so far been important in the fields of microelectronics and sensors [1]. Till date, intensive research in the field of magnetic and electric ordering and their coupling has been explored. There are various materials and coupling mechanisms that have been explored in this field with ME coupling [2], [3], [4].

$BaTiO_3$ (BTO) is a well-known ferroelectric material. Chemical modification of BTO can be done to impart ferromagnetism to the crystal structure and develop a probable magnetoelectric material [5]. Transition metal (TM) doping at Ti site is an effective method to achieve BTO-based magnetoelectric materials [6], [7]. $BaTiO_3$-$BiFeO_3$ [8], $BaTiO_3$-$LaFeO_3$ [9], $BaTiO_3$-$La_{0.7}Ba_{0.3}MnO_3$ [10], Mn and Co modified $BaTiO_3$-$BiFeO_3$ [11], [12], [13] etc. are some examples of solid solutions of BTO with magnetic ($ABO_3$) perovskite materials which demonstrate ME coupling. Apart from doping with magnetic ions and solid solutions, composites with ferromagnetic materials [14] also reveals ME coupling. Core shell structures with magnetic materials [15] and heterojunction thin films of BTO with ferromagnetic films [16], [17] are also capable of exhibiting ME coupling. Excellent ME coupling was revealed in most of these materials leading to the exploration of rich physics and utilization as industrial applications. BTO can be made multiferroic by doping with a TM ion at the Ti -site to induce magnetism in BTO and simultaneous substitution of Ba by elements such as Ca, Bi, Sr, La at the A-site to retain the ferroelectricity of the structure [6][18]. This will show magnetic orderings while retaining the ferroelectricity of $BaTiO_3$, leading to multiferroicity and magnetoelectric coupling.

Khedri et al. [19] investigated Ca-doped BTO and revealed the retainment of the tetragonal *P4mm* structure even up to 20% Ca doping. Hasan et al. [20] supported the experimental work with a theoretical study on the structural and physical properties of Ca-doped $BaTiO_3$. On the other hand, Tong et al. [21] experimentally revealed a ferromagnetic exchange mechanism in Mn-doped BTO that involves a spin-polarized hole trapped at Ti vacancies leading to effective exchange interactions between hole carriers and magnetic Mn ions. Second order Jahn Teller effect due to $Mn^{3+}$ ions at Ti sites favoring ferroelectricity and favors ferromagnetism due to its magnetic exchange interaction. Hence, ferromagnetism can be induced by retaining the ferroelectricity by both A-site Ca- doping and B-site $Mn^{3+}$-doping in BTO. Multiferroicity can happen due to the ferromagnetic exchange interactions of the d electrons in TM ions. Hence, an attempt to incorporate Ca and Mn at the A and B-site respectively has been made in this work to investigate a structure-correlated multiferroicity and probable magnetoelectric coupling in modified BTO.

Nayak et al. [22] studied the emergence of hexagonal *$P6_3/mmc$* with Cr doping at Ti site of BTO. Dang et al. [23] investigated the structural polymorphism from *P4mm* to *$P6_3/mmc$* with Mn doping in Ti site of BTO. Pal et al. [6] reported the role of hexagonal $P6_3/mmc$ space group in inducing ferromagnetism in Fe-doped and Mn-doped tetragonal BTO. They have also studied how to tune this hexagonal phase to zero in Fe and Mn doped BTO by simultaneous doping at A-site with various elements. Hence, the doping at A-site and B-site with suitable elements in BTO can effectively tune the simultaneous presence of tetragonal *P4mm* and hexagonal $P6_3/mmc$ structures. Such a dual phase presence can not only tune the multiferroicity but also couple these phases to impart ME coupling. Mn-doping at B-site introduces magnetic exchange interaction but leads to the emergence of paraelectric $P6_3/mmc$ space group but simultaneous doping of Ca at A-site inhibit the entire conversion of tetragonal BTO to hexagonal BTO and helps for retaining ferroelectricity. Structure correlation of the ferroelectric, ferromagnetic properties are investigated in detail and coupling of the electric and magnetic orders are measured and quantified as the magnetoelectric voltage generated in the material. The underlying physics of magnetoelectric properties exhibited by these materials has been explored and reported. A complete structural correlation linked with the physical properties has been reported in this work.

**Experimental details**

Ca and Mn are doped in $BaTiO_3$ with chemical formula, $Ba_{1-x}Ca_xTi_{1-y}Mn_yO_3$. The samples were named BTO (x = y= 0), BCTMO1 (x = y= 0.03), BCTMO2 (x = y= 0.06), BCTMO3 (x = y= 0.09). A sol-gel method was adopted to synthesize polycrystalline powders of the compositions. [ Detailed synthesis procedure has been mentioned in Supplementary [SE1]. The calcination has been done at 1200 °C for 3 hours and were pressed into pellets then sintered at 1330 °C for 4 hours. X-Ray Diffraction (XRD) technique was used to verify the phase of the all compositions using a Bruker D2 – Phaser X-Ray Diffractometer. The data were recorded at a sweep rate of $0.6^0$ min$^{-1}$ using Cu Kα radiation (λ =1.54 Å) at 30kV and 10mA. Raman spectroscopy has been done using a Horiba-made LabRAM HR Raman spectrometer (spectral resolution of 0.9 cm$^{-1}$) at laser of wavelength 632.8 nm. X-ray photoelectron spectroscopy (XPS) was performed using a Thermo-Scientific Escalab-250-Xi XPS spectrometer to evaluate the chemical composition, bonding environment, and the oxidation states of the elements present in the samples. Monochromatic Al-*Kα* X-rays source was used to investigate the materials. A flood gun was used to irradiate the samples and neutralize the charging effects. The O1s, Ba3d, Ti2p, Ca2p and Mn2p high-resolution XPS spectra were recorded at a pass energy of 20 eV. The base pressure in the chamber was originally 1 × 10⁻¹⁰ mbar. The EXAFS measurement of the K-edges Ti (4966 eV) and Mn (6539 eV), X-ray absorption spectra were recorded at Scanning EXAFS Beamline (BL-09) of INDUS-2 Synchrotron Source, Raja Ramanna Centre for Advanced Technology (RRCAT), Indore, India. The beamline equipped with Si (111) based double crystal monochromator for energy selection and meridional cylindrical mirror (Rh/Pt coated) for collimation. The data was collected when the synchrotron source 2.5 GeV ring was operated at 120 mA injection current in transmission mode for Ti K-edge and in fluorescence mode for Mn K-edge at room temperature. An ion chambers were filled with $N_2$, He and Ar during calibration and measurement. The second crystal of the monochromator was 60% detuned during the data collection to suppress the higher harmonic components. The energy calibration was performed using Ti and Mn foils as a reference. A ferroelectric (P-E) Loop Tracer instrument (M/S Radiant Instruments, USA) installed in RRCAT, Indore, India was used to investigate the P-E loop analysis of the samples. Positive Up and Negative Down (PUND) measurement has been done on (M/S Radiant Instruments, USA) installed in UGC DAE, Indore, India. The magnetic properties were performed in a Physical Property Measurement System (PPMS, Quantum Design) using a vibrating sample magnetometer (VSM) installed at State university at New York, Buffalo state university, USA. Magnetoelectric coupling coefficient has been measured using Radiant Technologies Precision Multiferroic tester installed at Thapar University, Punjab, India. The obtained P-H loop was analyzed and fitted with Radiant's Precision Line of Test Systems driven by Vision Software.

**RESULTS AND ANALYSIS**

**X-Ray diffraction studies**

XRD data of the BTO sample confirms a tetragonal *P4mm* structure. All BCTMO samples have a mixed phase of tetragonal *P4mm* and hexagonal *P6₃/mmc* structure of $BaTiO_3$ [Figure1a]. There is no other impurity peaks observed and well-defined sharp peaks confirm the crystalline $BaTiO_3$. Figure 1b shows a zoomed image of the highest intensity tetragonal peak, T(110). The existence of an additional peak corresponding to highest intensity hexagonal peak H(104) as a shoulder peak to T(110) peak start to appear from BCTMO1 onwards and its intensity gradually increases as the doping increases [Figure 1b]. This indicates the increase of the hexagonal phase percentage in the samples. The tetragonality of the samples are indicated by the splitting of T(002) and T(200) peaks as shown in Figure 1c. The splitting of these peaks for all samples indicates the tetragonal phase still coexists with the hexagonal phase of BaTiO$_3$.

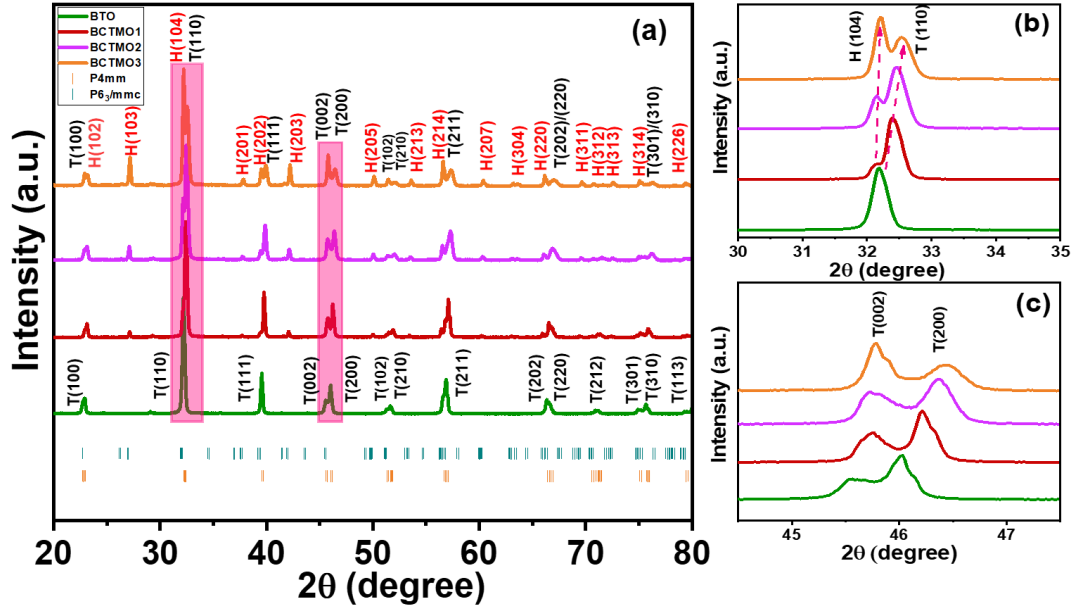

**Figure 1:** (a) XRD data of all the samples with cif file of P4mm and P6$_3$/mmc space group, (b)Zoomed image of the variation of the intensity of the hexagonal peak H(104) and tetragonal peak T(110) with composition, (c) Zoomed image of the splitting of tetragonal T(002) and T(200) peaks.

The structure was analyzed by Rietveld refinement using Fullprof software [Figure 2a]. The XRD data has been refined with the tetragonal *P4mm* standard CIF file [25] and Hexagonal *P6$_3$/mmc* standard CIF file [26] obtained from Crystallography Open Database [Figure 2b]. The refined files of all the samples are included in supplementary file Figure S1. The phase percentage, lattice parameters, bond lengths and bond angles are extracted from the refined CIF files of the samples. The tetragonal phase reduces from 100% (BTO) to 88% (BCTMO1), 71% (BCTMO2) and 45% (BCTMO3). Hence, the hexagonal phase increases from 0% (BTO) to 12% (BCTMO1), 29% (BCTMO2) and 55% (BCTMO3) [Figure 2c]. Therefore, a major tetragonal *P4mm* phase was observed in BTO, BCTMO1, BCTMO2, while a major hexagonal *P6$_3$/mmc* phase was observed in BCTMO3. The presence of d electrons in transition metals (TM) when doped at the B-site of BTO, is reported to trigger hexagonal phase (*P6$_3$/mmc*) [6]. Note that the hexagonal phase induced in the doped samples represent the hexagonal phase of BaTiO$_3$ itself and not that of the BaMnO$_3$ [27] and CaMnO$_3$ [28]. This can be confirmed from the 2θ values and the intensity profile of the two structures. Hence, the incorporation of Mn in BaTiO$_3$ is solely responsible for the drastic conversion of the structure from a *P4mm* to *P6$_3$/mmc* and incorporation of Ca in BaTiO$_3$.

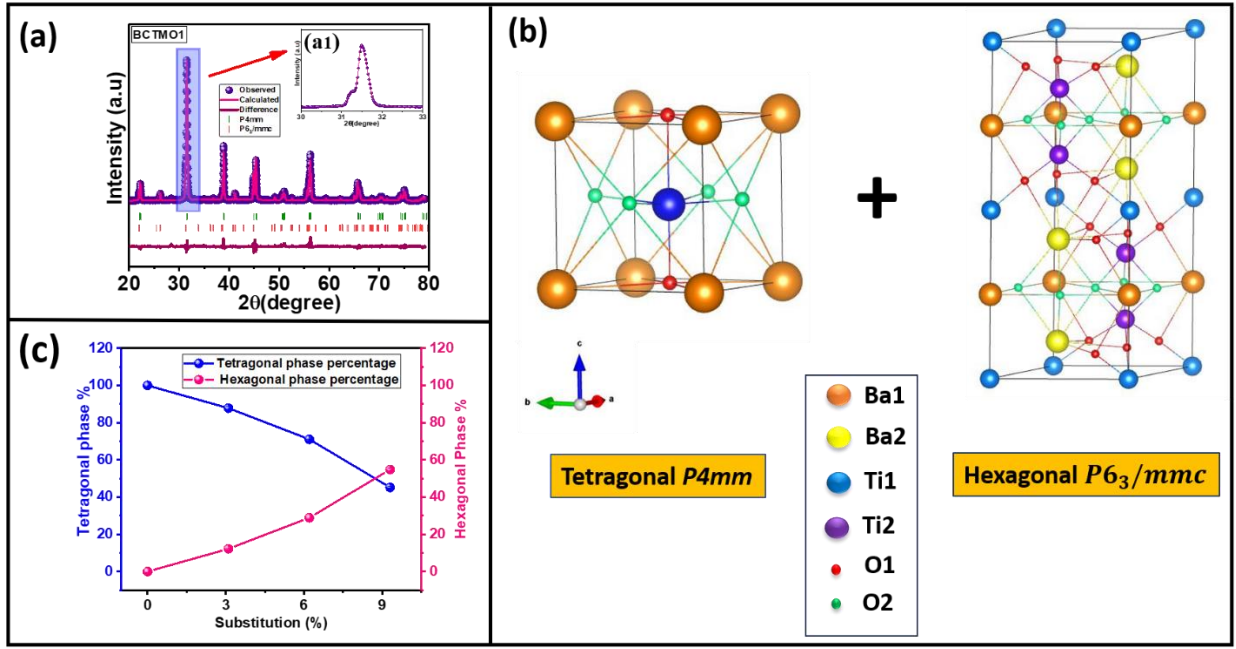

**Figure 2:** (a) Rietveld refinement of the BCTMO1 sample with P4mm and P6$_3$/mmc space group, inset: (a1) zoomed image of the T(110) and H(104) peak of the refinement using both phases (b) Variation of the phase percentage of the tetragonal and hexagonal phase with composition, (c)Schematic of the crystal structure of Tetragonal P4mm and hexagonal P6$_3$/mmc using VESTA software.

In an ideal BaTiO$_3$, Ti will be in Ti$^{4+}$ oxidation state. However, in an experimental sample to achieve an exclusive Ti$^{4+}$ valence state is a challenge. Hence, practically in most reported samples a mixture of Ti$^{3+}$ and Ti$^{4+}$ is obtained thereby, inducing the defects like oxygen vacancies (O$_V$). Similarly, Mn doping can lead to the possibility of Mn$^{3+}$ and Mn$^{4+}$ states with d$^4$ and d$^3$ electrons and can trigger structural distortion leads to transformation to a *P6$_3$/mmc* phase, inciting an instability of the tetragonal polar phase. This can be due to the changes in the TM3d - O2p hybridization or pseudo Jahn Teller distortion induced by Mn$^{3+}$, which is absent in Ti$^{4+}$ ions.

The non-centro symmetry of the Ti atom (blue color atom Figure 3a) in the TiO$_6$ octahedra of the *P4mm* structure is responsible for the ferroelectricity in BTO. However, in P6$_3$/mmc structure, the unit cell consists of eight corner shared Ti atoms and four edge shared Ti atoms of first kind, Ti1 (blue color atoms in Figure 3b), and four internal Ti atoms of second type, Ti2 (Violet color atoms in Figure 3b). Both Ti1, Ti2 atoms form two types TiO$_6$ octahedra in terms of symmetry. Ti1forms TiO$_6$ octahedra which are symmetric in nature with six equal Ti-O bond lengths ~1.9905 Å. On the other hand, two neighboring octahedra of Ti2 atoms share a face of three O atoms, represented as O2 (pale green atoms in Figure 3b). Such a conjunct pair of face shared octahedra are generally together called a dimer. These internal Ti2 atoms in the dimer form $TiO_6$ octahedra which are distorted. The arrangement of these distorted $TiO_6$ octahedra is a result of the transformation from *P4mm* to *P6₃/mmc* structure.

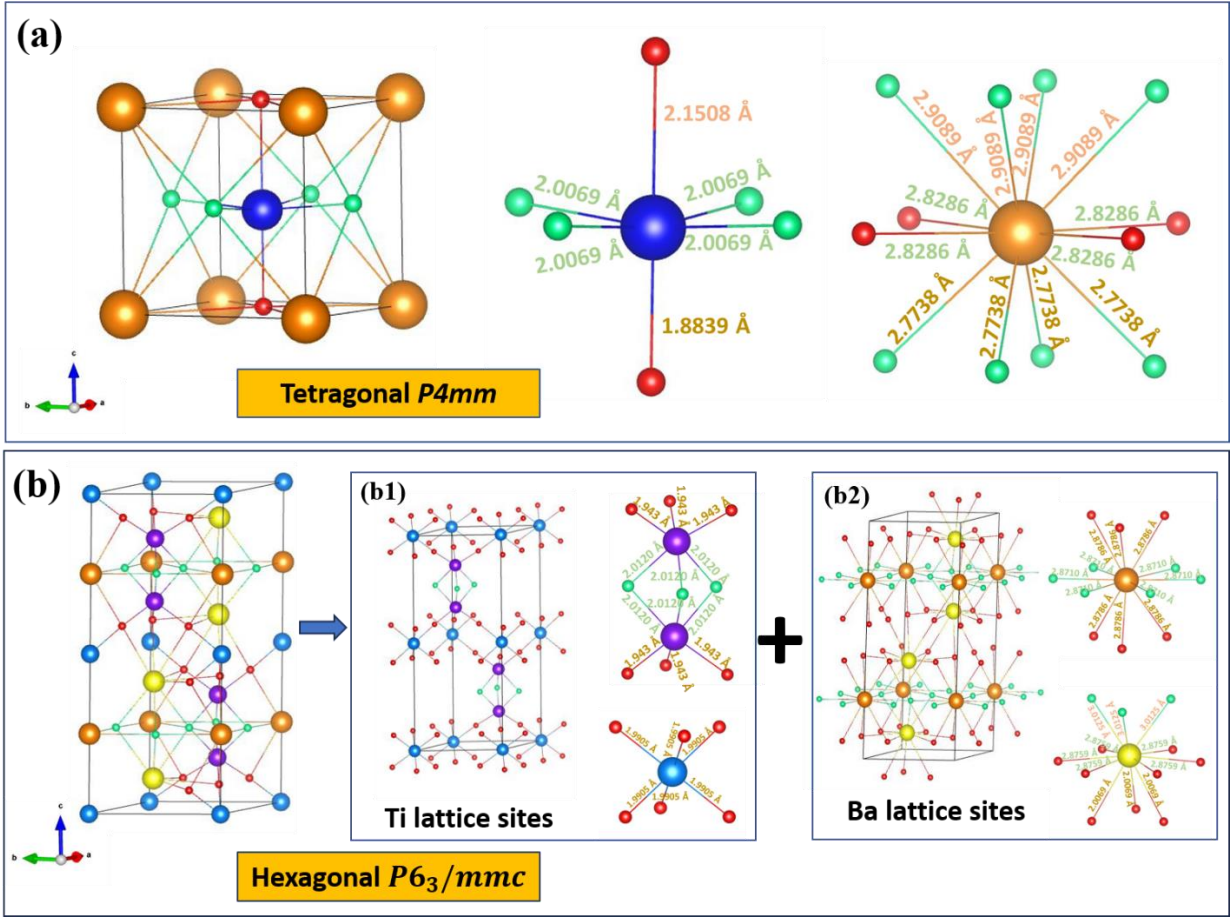

**Figure 3:** (a) Schematic of the *P4mm* structure and different types Ti-O and Ba-O bonds, (b) Schematic of the *P6₃/mmc* structure, (b1) Schematic of the Ti and O lattice in *P6₃/mmc* structure with two types of Ti-O bonds (b2) Schematic of the Ba and O lattice in *P6₃/mmc* structure with two types of Ba-O bonds.

The distance between the Ti2 ions inside the dimer is close to ~2.8 Å for the hexagonal $BaTiO_3$ [26]. The repulsion between the two Ti2 atoms results in a longer Ti-O bond near the $O_3$ face of dimer, resulting in six Ti-O bonds ~1.992 Å and three shorter Ti-O bonds ~1.958 Å that are part of each Ti2 atom away from the $O_3$ face. Hence, the two $TiO_6$ octahedra in the dimer are distorted having three shorter and three larger Ti-O bonds [Figure 3b]. The relative displacement of the Ti2 ions w.r.t. the O-cage of the distorted TiO$_6$ octahedra in the hexagonal BaTiO$_3$ are on the opposite sides thereby negating the possibility of a ferroelectricity due to Ti ions [Figure 4a]. With the initial Mn incorporation, the average Ti-Ti separation changes to 2.78 Å in BCTMO1, 2.67 Å in BCTMO2, and 2.423 Å in BCTMO3 [Supplementary Figure S2]. Hence, it appears that the Ti atoms are moving closer to each other. Note that although it is difficult to comment on the O position from the XRD results, a trend of reducing TiO$_6$ octahedral volume was observed with increase in doping concentration. Hence, there is a serious indication of a strong Ti-Ti interaction. The reason behind such an increase in strength of this interaction needs to be explored and may be related to the modifications in the nature of the Ti-site, i.e, the nature of the valence state of the Ti atom or Mn atom.

Apart from the TiO$_6$ octahedra, the *P6$_3$/mmc* structure consists of a repeating pattern of three Ba-O planes [(Ba2-O1), (Ba1-O2), (Ba2-O1)] as shown in Figure 3b. In this the consecutive Ba2 atoms are displaced from the O1 plane in opposite direction along c axis [Figure 4 b]. Similarly, when considering the two consecutive Ba1-O2 plane of one repeating pattern with Ba1-O2 plane of consecutive repeating pattern, there is minimal separation of Ba1 from the O2 plane in opposite direction along b axis [Figure 4b]. The displacement of Ba2 from Ba2-O1 plane along c direction is larger than that of displacement of Ba1 from Ba1-O2 along b direction. The displacement of two consecutive Ba1 atoms and two consecutive Ba2 atoms are displaced from the corresponding Ba-O plane in opposite direction and hence negate the possibility of ferroelectricity. The separation of the Ba2 atom from Ba2-O1 decreases with substitution from 0.25 Å (BCTMO1) to 0.21 Å (BCTMO2) and 0.16 Å (BCTMO3). On the other hand, the separation of the Ba1 atom from Ba1-O2 plane negligibly decreases with substitution from 0.18368 Å (BCTMO1) to 0.18323 Å (BCTMO2) and 0.18304 Å (BCTMO3), which is technically unreliable. However, the trend of reduction of the separation is there. Hence, without commenting strongly on such calculations one can visualize the effect as a probability.

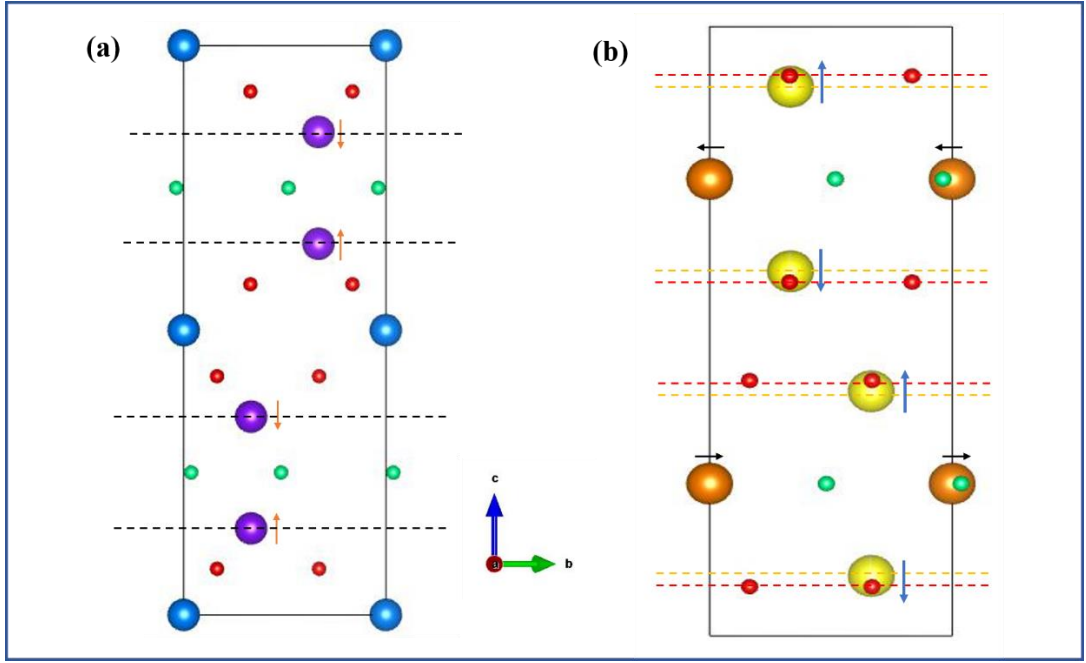

**Figure 4:** (a) Displacement of Ti2 atoms in opposite direction from the centrosymmetric position of the $Ti_2O_9$ dimers, (b) Displacement of consecutive Ba atoms in opposite direction from the Ba-O plane along c direction (represented by blue arrows) and along b direction (represented by black arrows).

Hence, it is observed that the introduction of Mn enables the TM ions in the dimer to come closer to each other whereas the introduction of Ca enables reduction of the separations of the Ba(Ca) and O planes. It has been observed from literature that Mn doping promotes the phase transition from *P4mm* to *P6₃/mmc* and with higher doping to *R-3c* [23] while Ca doping retains the *P4mm* phase for as much as 20% doping [19]. Hence, a possibility of a phase transition towards *P6₃/mmc* with doping of Mn can be counteracted by a simultaneous doping of Ca which wants to retain the *P4mm* structure. This struggle between the two mechanisms may have generated a mixed phase of the doped samples.

The 'a' and 'c' lattice parameters of the both phases also reduce with substitution [Figure 5a]. Hence, the volume of the unit cell seems to be decreasing with increase in doping concentration. This happens for both the *P6₃/mmc* and *P4mm* space group. As a result of this volume contraction the c/a ratio of the *P4mm* phase also reduces continuously [Figure 5b]. The tetragonal lattice parameter 'a' and 'c' negligibly changes from BTO to BCTMO1 then shows a gradual decrease till BCTMO3. Similarly, 'a' and 'c' lattice parameters of hexagonal phase show gradual decrement from BCTMO1 to BCTMO3. Changes in the lattice parameters can be explained through ionic radii of the dopants in A-site and B-site. $Ba^{2+}$ [1.75 Å (XII)] is replaced by smaller ionic radii $Ca^{2+}$ [1.48 Å (XII)] at the A-site of BTO. However, Mn can have the possibility of $Mn^{3+}$ [0.785 Å (VI)- high spin, 0.72 Å (VI) low spin] and $Mn^{4+}$ [0.67 Å (VI) oxidation state in $Ti^{4+}$ [0.745 Å (VI)] $Ti^{3+}$ [0.81 Å (VI)] site [29]. The doping of lower ionic radii $Mn^{4+}$ in $Ti^{4+}$ site reduces the lattice parameters. Whereas, $Mn^{3+}$ doping will enhance the lattice parameters. Though there is reduction in lattice parameters due to A- site modification, the lattice parameters are not changed drastically due to the competition of $Mn^{3+}$ and $Mn^{4+}$ oxidation states in Ti site. There are only minor changes in the c/a ratio indicating the retention of tetragonality even for the highest doped samples.

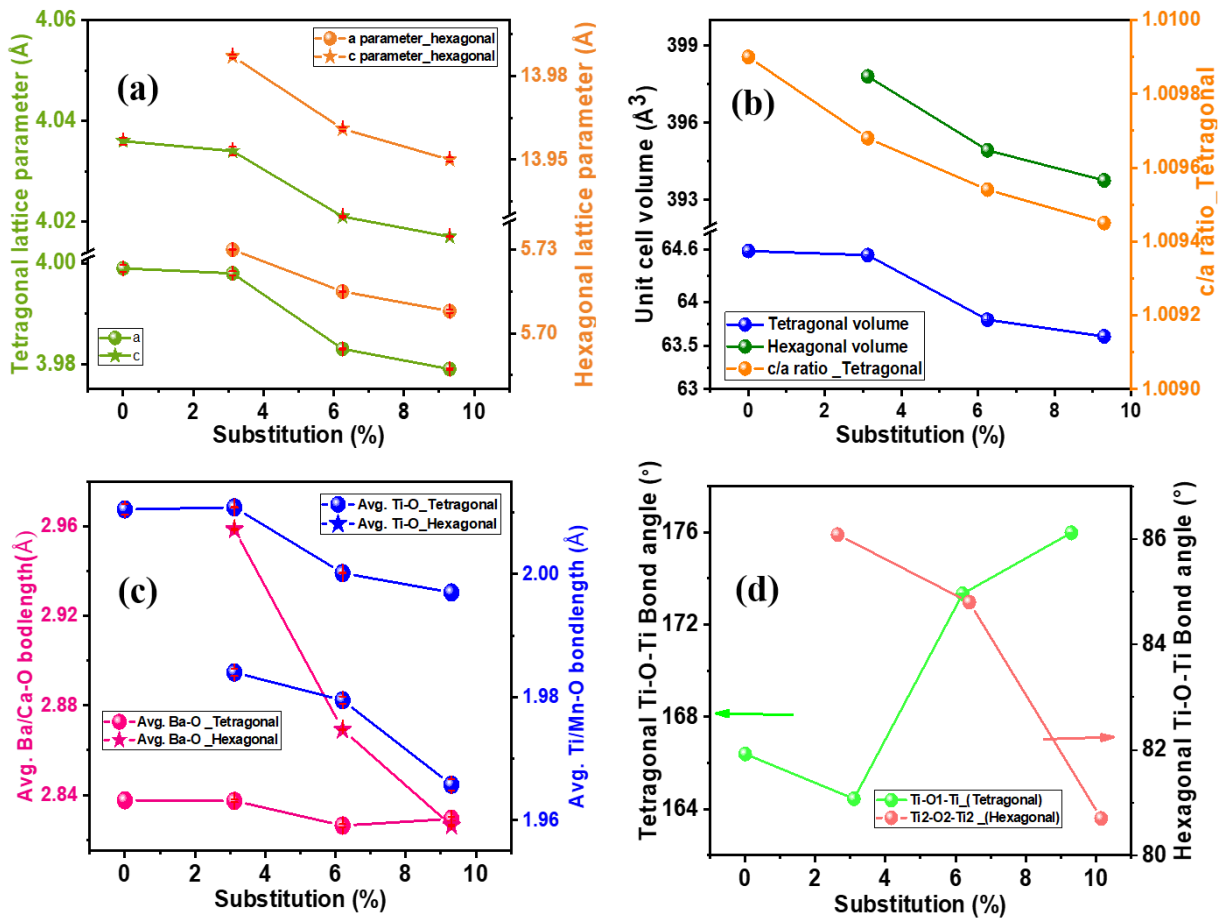

**Figure 5:** (a) Variation of tetragonal and hexagonal lattice parameters with composition, (b) Variation of average Ba/Ca-O and average Ti/Mn-O bond length with composition, (c) Variation of tetragonal and hexagonal volume and tetragonal c/a ratio with composition, (d) Variation of tetragonal and hexagonal Ti-O-Ti bond angle with composition.

The dopant ionic radii also influence the average Ti/Mn-O and Ba/Ca-O bond lengths. There is a gradual reduction in these average bond lengths [Figure 5c]. Variations in the 180° Ti-O-Ti bond angle in tetragonal lattice are analyzed and plotted [Figure 5d]. There is a slight decrease of the bond angle for BTO to BCTMO1 then it gradually increases. For the hexagonal phase variation in 90° Ti2-O1-Ti2 bond angle shows a decrease in trend. The changes in the bond angle in both the phases indicated the variations in the overlapping of the atomic orbitals and lead to changes in the magnetic exchange interactions and variation in the magnetic properties accordingly. The probable changes in the magnetic exchange with structure will be discussed later.

**Raman analysis**

The Raman spectrum of all the samples reveal E(TO) phonon mode at 40 cm$^{-1}$, $A_1$(TO) mode at 280 cm$^{-1}$, $B_1$/E(LO+TO) mode at 305 cm$^{-1}$, $A_1$(TO)/(E(TO) mode at 520 cm$^{-1}$, $A_1$(LO)/E(LO) mode at 720 cm$^{-1}$ corresponding to tetragonal *P4mm* structure [Figure 6a] [30]. As the doping percentage increases all the modes corresponding to tetragonal structure observed to be broaden. This is an indication of presence of different types of vibration due to the dopant induced variations of bonds. Some phonon modes corresponding to the hexagonal phase are present in the samples at ~ 100 cm$^{-1}$ [$A_{1g}$], ~420 cm$^{-1}$ [$E_{2g}$] and 640 cm$^{-1}$ [$A_{1g}$] [31]. The coexistence of the characteristic peaks of both tetragonal and hexagonal phases in the doped samples confirm the dual phase nature of the samples. Figure 6(b) and Figure 6(c) shows the variation in intensity and broadening of the tetragonal mode at ~305 cm$^{-1}$ and hexagonal mode at ~640 cm$^{-1}$ respectively. The reduction of the peak intensity of 305 cm$^{-1}$ indicates the reduction in the tetragonal phase percentage and the increment in the intensity of the ~640 cm$^{-1}$ indicates the increase in the hexagonal phase percentage [Figure 6d]. Due to the presence of a heavier Mn ion w.r.t Ti ion a probable redshift is observed for the characteristic peaks of both the phases [Figure 6e].

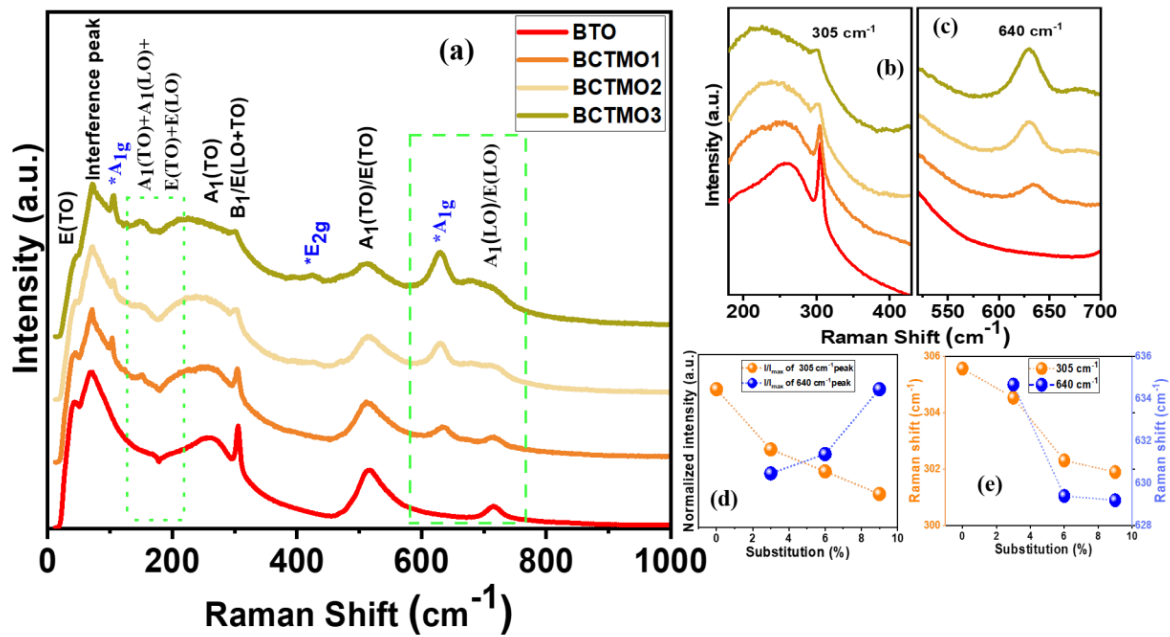

**Figure 6:** (a) Raman spectra of all samples, (b) Zoomed image of the variation of 305 cm$^{-1}$ tetragonal mode, (c) Zoomed image of the variation of 640 cm$^{-1}$ hexagonal mode and 720 cm$^{-1}$ tetragonal mode, (d) Variation of intensity of 305 cm$^{-1}$ and 640 cm$^{-1}$ mode with composition, (e) Variation of Raman shift of 305 cm$^{-1}$ and 640 cm$^{-1}$ mode with composition

**Valence state analysis**

XPS analysis of each element was performed to understand the oxidation states of various elements present in the material. The oxidation state is important to understand as the bonding of electrons with the neighboring ions are correlated to the structure and hence the physical properties. Due to spin orbit interaction Ba 3d, Ca 2p, Ti 2p, Mn 2p peaks are observed to display doublet features with two separated peaks. The area fraction of each ion can be calculated with general formula *$[A^{n+}]/([A^{n+}]+[A^{m+}])$*, where A is an ion having multiple oxidation states n+ and m+ present, with $[A^{n+}]$ and $[A^{m+}]$ being the area under the corresponding peaks in the XPS spectrum. This area fraction can be treated as the contribution of $A^{n+}$ content over $A^{m+}$. The area fractions of $Ti^{3+}$ over $Ti^{4+}$ ($f_{Ti3+}$), $Mn^{3+}$ over $Mn^{4+}$ ($f_{Mn3+}$) are calculated from the deconvoluted XPS spectra. The fractions $f_{Ti3+}$ and $f_{Mn3+}$ contain information about the loss of positive charge. It should be noted that Ba and Ca are predominantly in the +2-charge state without the possibility of any other state. Hence, the effective cationic valence state will be dictated by these fractions. To maintain the charge neutrality of the crystal structure an easy solution is to have an oxygen deficiency. Hence, these fractions $f_{Ti3+}$ and $f_{Mn3+}$ are counter informative to the $O_v$. Therefore, these fractions were analyzed with increasing amounts of doping.

**Figure 7:** Deconvoluted XPS spectra of all samples using XPS PEAK 4.1 software (a) Ti 2p, (b) Mn 2p (c) Variation of area fraction of overall

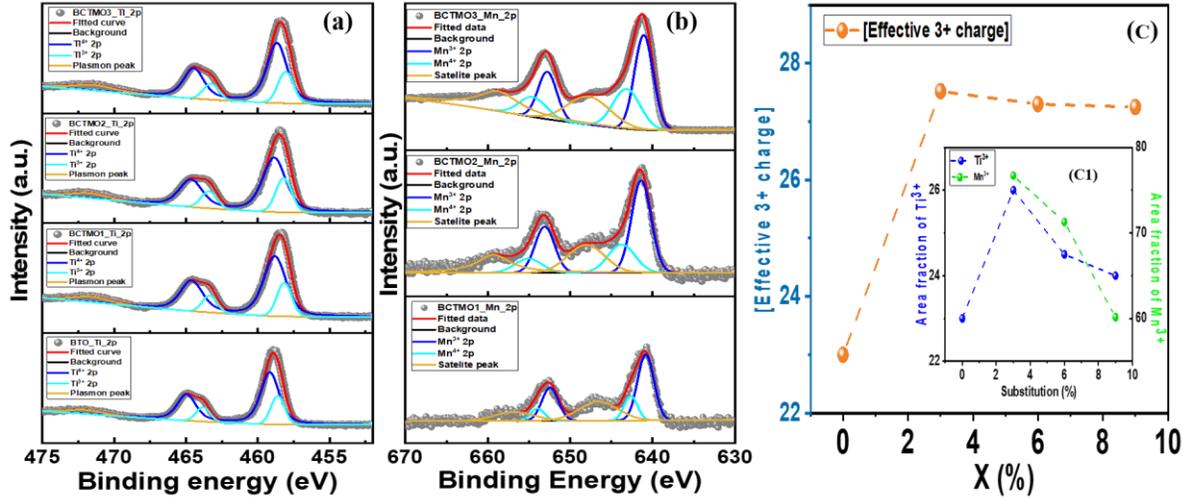

effective 3+ oxidation state with substitution, Inset (C1) shows the individual variation of area fraction of $Ti^{3+}$, $Mn^{3+}$ with substitution.

The Ti 2p XPS spectra in BaTiO$_3$ is a strong feature and reveals a doublet $Ti2p_{3/2}$ and $Ti2p_{1/2}$. In the BTO sample, this doublet is observed at ~459.16 eV ($Ti2p_{3/2}$) and at ~464.86 eV ($Ti2p_{1/2}$) with spin orbit splitting energy of 5.7 eV [Figure 7a]. These peaks are consistent with $Ti^{4+}$ oxidation state [30]. Peaks corresponding to $Ti^{3+}$ are obtained at ~458.61 eV($Ti2p_{3/2}$) and at ~463.91 eV($Ti2p_{1/2}$) with spin orbit splitting energy of 5.3 eV [32]. The spin orbit splitting energy was kept constant at 5.7 eV for $Ti^{4+}$ while 5.3 eV for $Ti^{3+}$ for the fitting of the doped samples. There is an increase of B.E.~0.5 eV for $Ti2p_{1/2}$ for all doped samples with respect to BTO indicating the modification of the Ti environment with doping. While, there are no significant changes in $f_{Ti3+}$ which was observed to be ~23%-26% for all samples [Figure 7c1] with a slight increase from BTO to BCTMO1 then slight decrease afterward. Mn 2p XPS spectra of the doped BCTMO samples [Figure 7b] also revealed two doublets corresponding to $Mn^{3+}$ at binding energies 641.05eV, 652.45 eV with spin-orbit splitting of 11.4 eV and doublets for $Mn^{4+}$ at 642.59 eV, 653.89 eV with spin-orbit splitting of 11.3 eV for BCTMO1 [33]. The spin-orbit splitting energy was kept constant for $Mn^{4+}$ and $Mn^{3+}$ for the fitting of the doped samples. Area fraction of $f_{Mn3+}$ and $f_{Mn4+}$ reveals the presence of both $Mn^{4+}$ and $Mn^{3+}$ ions in comparable ratio with a slight dominance of $Mn^{3+}$ over $Mn^{4+}$. The value of $f_{Mn3+}$ was observed to decrease from BCTMO1 to

BCTMO3 from 77% in BCTMO1 to 70% in BCTMO2 and 62% in BCTMO3 [Figure 7c1]. The variation of the $Ti^{3+}$ and $Mn^{3+}$ together will contribute to the effective charge of 3+ cations. Hence, the fraction of 3+ valence state ions can be calculated from the equation $f_{3+} = (1-x).f_{Ti3+} + (x).f_{Mn3+}$, where, $x$ is the doping fraction. A drastic increase from BTO to BCTMO1 was observed for $f_{3+}$. However, for BCTMO2 and BCTMO3 not much difference was observed thereafter [Figure 8c]. Hence, with doping the structure becomes more defective due to a growing amount of $f_{3+}$ that can also indicate the increase of $O_V$.

The O1s XPS spectra of BTO can be deconvoluted to three peaks [Figure 8a] corresponding to the Binding Energy (B.E.) of lattice oxygen ($O_I$) at ~ 529-530 eV, second peak feature ($O_{II}$) at ~531-532 eV, and third feature corresponding to adsorbed oxygen ($O_{III}$) at ~ 532.5-533.5 eV. These values are in accordance with literature [34]. The contribution $O_I$ is from the O atoms bonded to Ba, Ca, Ti and Mn ions in the samples. On the other hand, the origin of $O_{II}$ is controversial, and has been explained as a contribution from both chemisorbed oxygen species (e.g., hydroxyl groups) [35], [36] and oxygen vacancies $O_V$ [37], [38], [39]. A third feature $O_{III}$ is a contribution from the surface adsorbed carbonates ($CO_3^{2-}$) and water molecules ($H_2O$) [35]. In natural conditions due to the atmospheric exposure there is adsorption of water molecules that contributes to the $O_{III}$ feature.

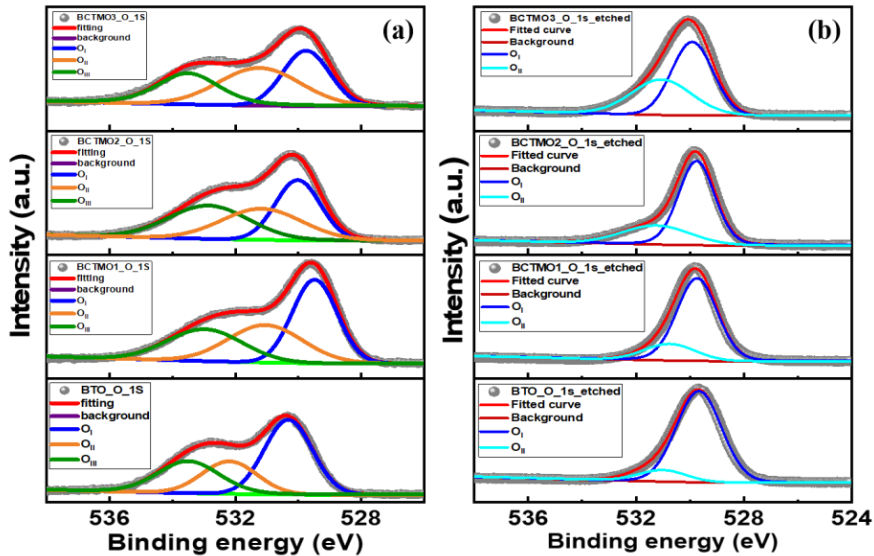

**Figure 8:** Deconvoluted XPS spectra of all samples using XPS PEAK 4.1 software (a) O 1s without etching (b) O 1s after etching.

There are numerous reports relating the $O_{II}$ feature (B.E. between 531 eV - 532 eV) to the presence of $O_V$. Wang et al. reported as evidence of generation of $O_V$ with increase of the $O_{II}$ peak intensity in $Fe_2O_3$ with increasing amount of annealing time in $N_2$ atmosphere, i.e., a reduced atmosphere revealing the connection between $O_{II}$ and $O_V$ [36]. However, this was an ex-situ study, which could have involved contributions from external contaminants. A recent in-situ XPS study revealed an un-changed spectra before and after reduction [40], [41], providing a strong support for the non-$O_V$ origin of $O_{II}$. Also, the absence of a O1s electron in the case of an $O_V$ has also been projected as a non $O_V$-origin of $O_{II}$ [36], [42]. Hence, it is difficult to judge from literature study, about the actual origin of $O_{II}$.

However, it is hard to believe that a crystal is so perfect that no $O_V$ will be present and the presence of $O_V$ will leave no mark on the XPS spectra. A loss of one O-atom should lead a change in the electronic charge distribution of the neighboring cations and should be evident from a dislocation of position of these cations [39]. Moreover, the coordination and most probably the valence state of the cations will differ due to the $O_V$ [37]. Such a dislocation and modifications will affect the local bonding of next nearest O-atoms and leads to changes in the binding energy. Hence, one should see an $O_V$ feature in the XPS spectra. However, whether the contribution is from $O_V$ or chemisorbed species is questionable. To try to answer this question, an etching of the surface layers on which chemisorption was possible was performed in this study for all samples. The O1s XPS data of the etched samples [Figure 8b] revealed a complete removal of the $O_{III}$ feature with a reduction of the $O_{II}$ feature. Note that etching is a process where inert atoms are bombarded on the sample surface to tear off the surface atoms from the surface layers and expose the inner bulk layers where chemisorption is not possible. A complete removal of the $O_{III}$ feature is strong evidence of the same. The reduction of the $O_{II}$ feature is also supporting the view that $O_{II}$ does have a component of chemisorption. However, the retainment of the feature in a reduced form ensures the contribution of other sources, such as $O_V$. Hence, the reduced $O_{II}$ feature is most probably a contribution to the $O_V$, in these samples.

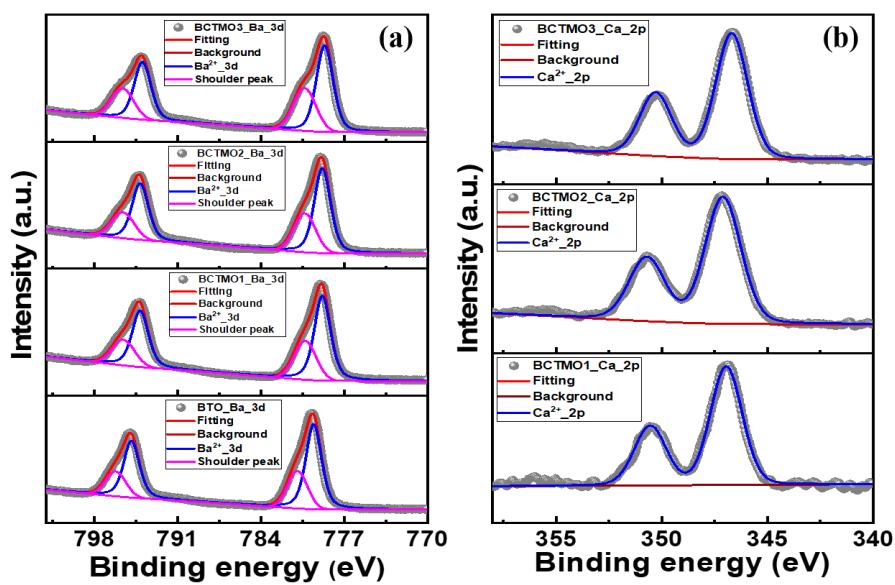

**Figure 9:** Deconvoluted XPS spectra of all samples using XPS PEAK 4.1 software (a) Ba 3d, (b) Ca 2p

The Ba XPS spectra [Figure 9a] reveal a doublet corresponding to $Ba^{2+}$ state $Ba3d_{5/2}$ at 779.6eV and $Ba3d_{3/2}$ at 795eV in BTO with spin orbit splitting energy of 4.6 eV. These peaks are resembling the reported features of the $Ba^{2+}$ state in the perovskite $BaTiO_3$ [43]. Shoulder peaks appear for both features corresponding to the different chemical state of the Ba atom on the surface [43]. All the doped samples are fitted with spin orbit splitting energy of 4.6 eV. There is a decrease of B.E. ~1 eV for $Ba3d_{5/2}$ for all doped samples with respect to BTO indicating the modification of the Ba environment with doping.

The Ca 2p XPS spectra of BCTMO samples [Figure 9b] shows doublet corresponding to $Ca^{2+}$ state at 347.63 eV ($2p_{3/2}$) and 351.23 eV($2p_{1/2}$) with a spin orbit splitting energy of 3.6 eV. These peaks are resembling the reported features of the $Ca^{2+}$ state in the perovskite $CaTiO_3$ [44]. Note that no prominent shoulder peaks are revealed for both features hinting at probably no different chemical state of the Ca atom on the surface. This indicates that Ca may not be prominently diffused to the surface of the particles or may not have participated in the process of adsorption. The positions of the $Ca3d_{5/2}$ and $Ca3d_{3/2}$ doublets have no major changes in the binding energies of the Ca ions in the doped samples, maintaining the spin orbit splitting energy of 3.6 eV, indicating a similarity in the Ca environment or bonding.

## XANES ANALYSIS

To further confirm the oxidation states, site occupancy and the local geometry of Ti and Mn ions, X-Ray absorption spectroscopy was performed. The special features and the absorption edge position in the XANES spectra gives the details of the symmetry of the orbitals, hybridization, and the oxidation states of the absorbing atom. The changes of the absorbing energy, indicated by the shifts in the XANES spectra, are indications of probable changes of the oxidation states of atom. The standard normalization and background subtraction procedures were executed using the ATHENA software version 0.9.26 to obtain normalized XANES spectra [45]

The Ti-K edges of the samples [Figure 10a] reveal sharp absorption features with a main peak at ~4987 eV (feature E) corresponding to the Ti 1s–4p transition and indicating the density of states of the unoccupied Ti 4p states [46]. This feature carries the information of the oxidation state of Ti in the crystal structure. The positions of this feature for the present samples are nominally shifted from ~ 4987 eV to lower energies with doping. This indicates a major $Ti^{4+}$ state in these samples. The main absorption peak intensity increases with doping. This is an indication of reduction in the non-centrosymmetric location of the Ti ion [47]. The non-centrosymmetricity introduces broadening of the 4p electronic states due to different hybridizations with the neighboring oxygen [46]. Hence, an increase of centrosymmetric nature is observed with doping. The nominal shift of the Ti-K absorption edge to lower energy value is observed mainly for BCTMO3 [Inset figure 10(b1)]. This indicates a decrease of oxidation state from $Ti^{4+}$ to $Ti^{3+}$.

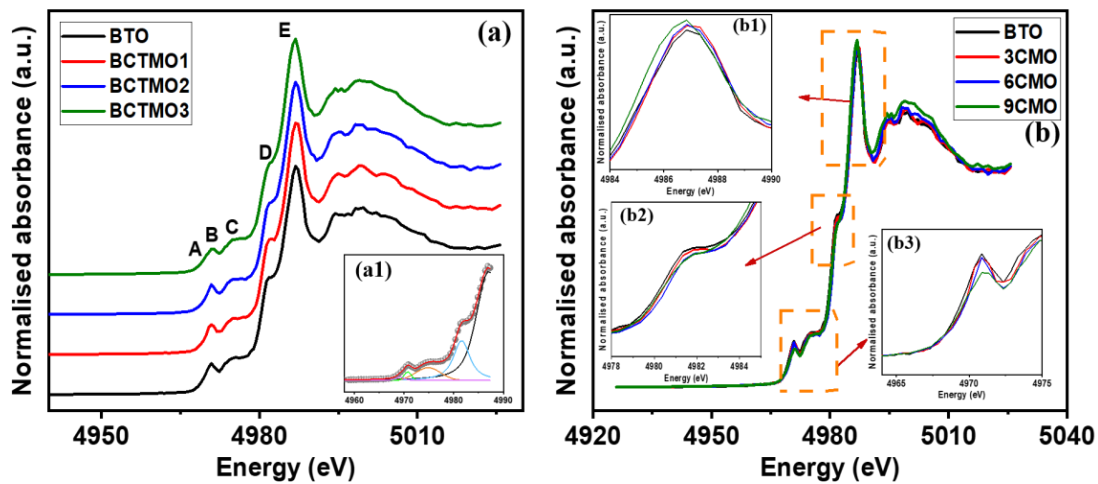

**Figure 9:** (a) XANES data of Ti K-edge with absorption peak E and all the pre-edge feature marked as A, B, C, D. (b) Variation of the features with samples, inset (b1) Zoomed image of the absorption peak E, inset (b2) Zoomed image of the pre-edge feature C, inset (b3) Zoomed image of the pre-edge feature B.

Small features at a lower energy (4965–4985 eV) represent the Ti 3d pre-edge features. There are four features marked as A (~4968 eV), B (~4971 eV), C (~4975 eV) and D (~4982 eV) [Figure 9a][48], [49] . The pre-edge region of the XANES spectra can be deconvoluted into four peaks in Athena software [Inset Figure 10(a1), Fitted data of all samples have been given in supplementary figure S4]. In the case of an octahedral symmetry, the five-fold degenerate 3d orbitals split into three-fold ($t_{2g}$) and two-fold ($e_g$) degenerate states. These degenerate states manifest as two pre-edge features: a minor hump-like feature, A for a Ti1s→Ti3d($t_{2g}$) quadrupolar transition and a sharp peak, B for 1s→Ti3d($e_g$)-O2p hybridized states which is mostly a dipolar contribution. The feature A is forbidden in a dipole approximation and is allowed in quadrupole approximation making it a very weak pre-edge feature [46]. This feature is so weak that analysis of this peak is irrelevant. As the hybridization between the Ti3d $e_g$ orbitals and O2p orbitals becomes more pronounced the local distortion in the $TiO_6$ octahedron increases resulting in a relatively large dipole component, hence, the intensity of the $e_g$ peak, B feature increases. The intensity of B and the mean-square displacement of Ti ion from the center of symmetric position of the $TiO_6$ octahedron are proportional, thereby B feature representing asymmetry of the $TiO_6$ octahedra and is absent if the $TiO_6$ octahedra is centrosymmetric [50], [51]. Note that this B feature is present in all samples indicating a non-centrosymmetric $TiO_6$ octahedra in all the samples [Inset figure 10(b3)]. This is a clear indication of the retention of the tetragonal phase with doping in all the samples. Intensity of feature B gradually decreases in intensity with doping, hinting at a gradual transformation towards a more centrosymmetric $TiO_6$ octahedra [Figure 11a]. This information matches the XRD results [Figure 11b]. It is interesting to note that in this mixed phase material, the hexagonal phase is composed of both centrosymmetric and non-centrosymmetric octahedra. The non-centrosymmetric $TiO_6$ octahedra form the dimers. From the XRD studies, these octahedra were observed to continuously evolve with doping with respect to the position of the Ti atoms. The Ti atoms were found to move closer to each other. But this distortion was found to be very less compared to the octahedral distortion in the tetragonal phase.

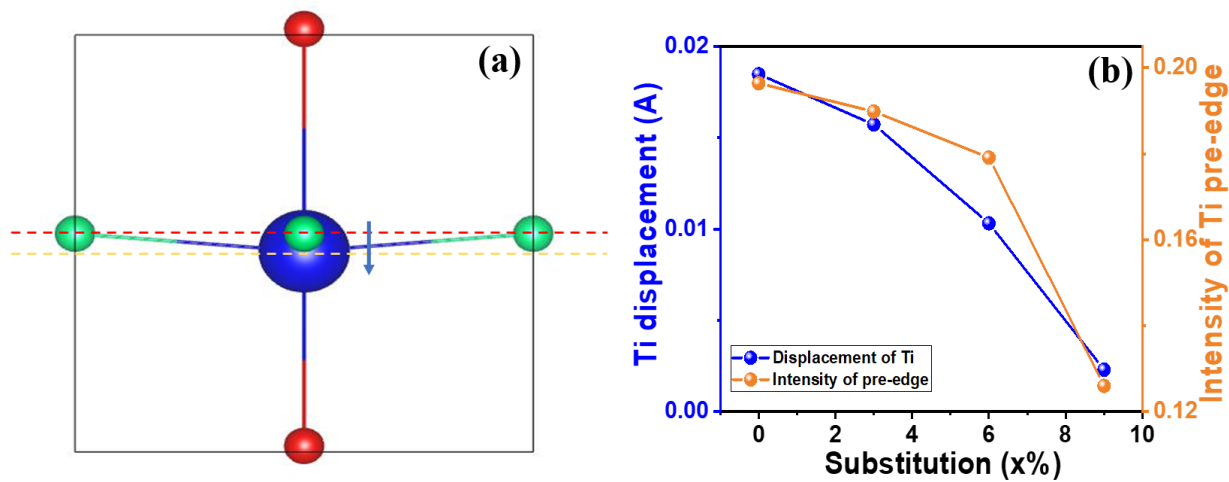

**Figure 11:** (a) Schematic of the displacement of Ti from centrosymmetric position in *P4mm* tetragonal structure of BaTiO$_3$, (b) Variation of Ti displacement from centrosymmetric position of BaTiO$_3$ (*P4mm*) and the variation of intensity of the Ti pre-edge feature B with composition.

Feature C (~4975 eV) appears due to the interaction with the neighboring Ti atoms and has a dipolar character [49]. There is no trend in the intensity profile of this feature indicating a dynamic change of Ti-O-Ti bond angles and Ti-O bond lengths. The shoulder structure D (~4982 eV) is a characteristic of tetragonal BaTiO$_3$ [Inset figure 10(b2)] [46]. The identification of the electronic states responsible for this shoulder feature is not been identified yet, but there are some conclusions that have been reported with the shoulder structure indicating the contribution of the A-site and B site ions in the ferroelectric perovskite. This reflects the electronic hybridization between Ti and Ba in the perovskite. A theoretical background for this claim based on the Born effective charges of each constituent atom is reported by Ghosez et al.[52] in their work. The modification of Ba site and Ti site of BaTiO$_3$ induce changes in the intensity of the shoulder peak indicating it is correlated with the electronic hybridization between A- site and B-site via the O atoms. There is a gradual decrease in the intensity of the peak D indicates the gradual decrease in the A-B bond with Ca an Mn doping in Ba and Ti site respectively.

The Mn-K edge XANES absorption peak of $Mn^{3+}$ is observed at ~6555 eV while $Mn^{4+}$ appears at ~6562 eV [53]. The normalized Mn-K edge in fluorescence mode for all samples [Figure 12a] is observed at ~6558 eV which is in between the absorption peaks of $Mn^{3+}$ and $Mn^{4+}$ [Inset Figure 12(b1)]. Pre-edge features corresponding to Mn 1s → 3d ($t_{2g}$) and 3d ($e_g$) states appear as a broad feature in the region ~6542 eV. These pre-edge features are not resolved experimentally due to core-hole lifetime broadening [50]. The intensity of the pre-edge feature for $Mn^{4+}$ is generally more than Mn$^{3+}$ [54]. The gradual evolution of the pre-edge intensity with doping indicates the relative increase of Mn$^{4+}$ over Mn$^{4+}$ with doping as obtained in XPS. Comparing the pre-edge intensities of Mn and Ti, a very feeble pre-edge intensities contribution is observed for Mn as compared to Ti for all samples [Figure 12 (b2)]. The MnO$_6$ octahedra, therefore, is expected to be relatively less distorted than TiO$_6$ octahedra. The XANES feature obtained can also indicate the occupancy of Mn at A site or B site. From the first-principles calculations, based on energetics, that Mn will predominantly occupy the B site in all equilibrium growth conditions rather than going to A site [55]. Note that the main Mn-K edge is a sharp absorption peak without any pre edge feature peak indicates a simultaneous replacement of Ba and Ti by Mn [56]. However, the similar XANES features of Ti K-edge and Mn K-edge with pre-edge features confirm the B site occupancy of Mn in BTO.

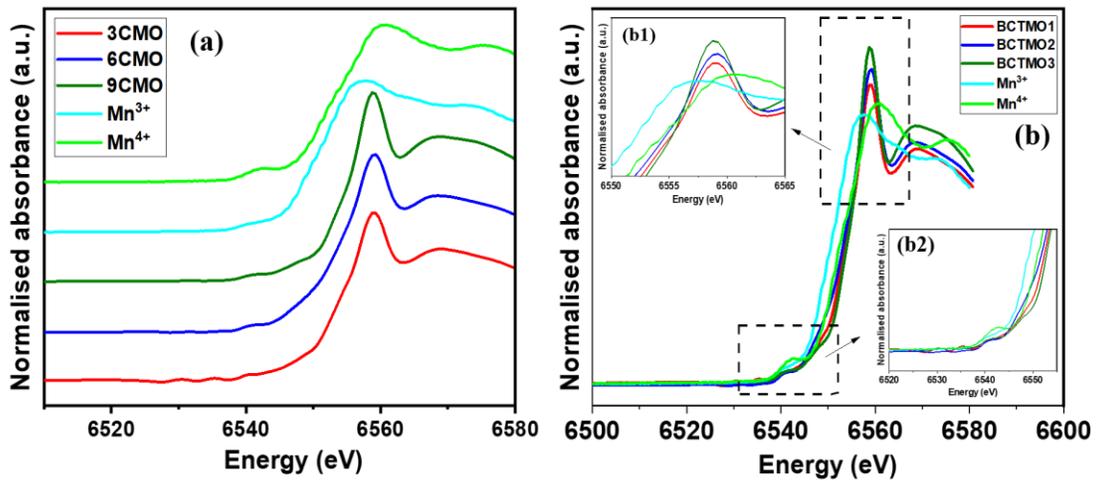

**Figure 12:** (a) XANES data of Mn K edge, (b) Variation of the features with samples, inset (b1) shows the zoomed images of the absorption peak, inset (b2) shows the zoomed image of the pre-edge feature.

**Ferroelectric properties**

The P-E hysteresis loop is the defining characteristic of ferroelectricity in materials. At zero field, ferroelectric BTO possesses randomly oriented ferroelectric domains. This allows a zero net polarization. However, the mode of experiment does not allow to visualize this zero-polarization state due to the application of an ac electric field of frequency 20 Hz [Figure 13]. A proper P-E loop is observed for BTO with saturation polarization of 12 μC/cm$^2$. This reveals the formation and alignment of the domains through domain wall movements with the application of field. Doping of Ca and Mn in BTO causes reduction in the remnant polarization ($P_r$) and saturation polarization ($P_s$). Since the addition of Mn promotes the formation of the hexagonal phase which is non-ferroelectric phase is a cause for this reduction of $P_r$ and $P_s$. In addition to this grain size, domain wall density and oxygen vacancies can also have significant influence on the ferroelectric properties of these materials [57].

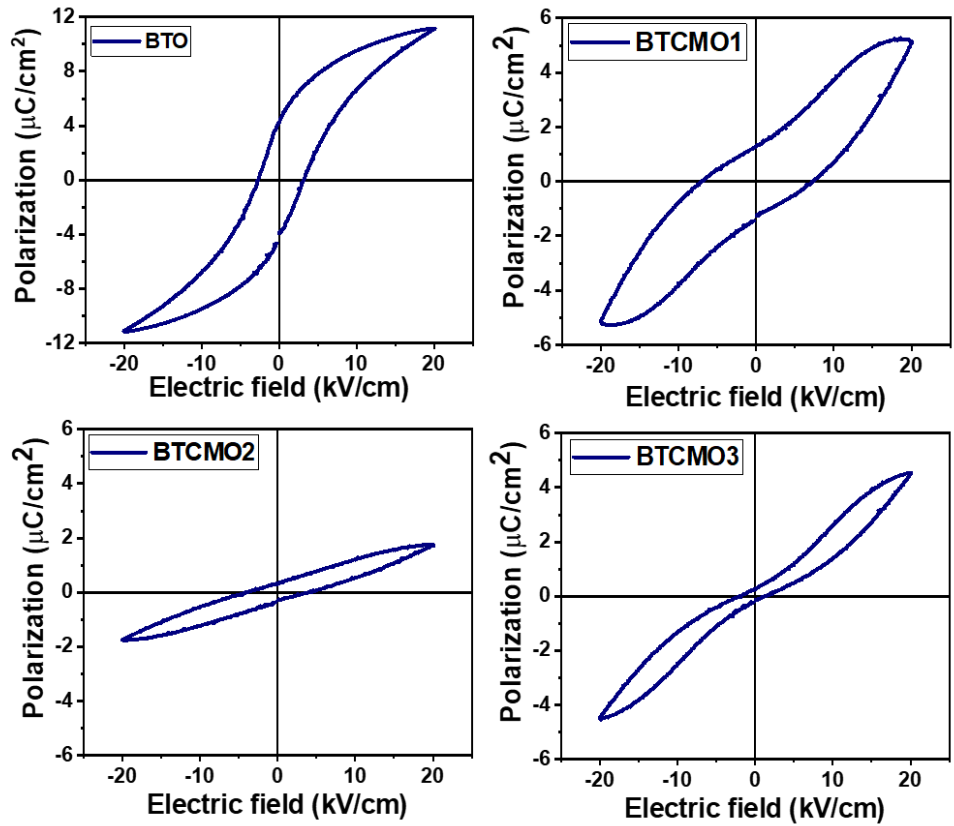

**Figure 13:** Room temperature P-E loop of all samples.

For BCTMO samples cigar-shaped P-E loops are observed. These loops do not get saturated, indicating the lossy nature of the samples [58]. Pinched loops are observed for all the doped samples. For perovskite ceramics, pinning of the charged defects dipoles on domain wall motion have been reported. Such pinning of defect dipoles can often lead to pinching of the P-E loops [59]. According to the symmetry conforming principle, the defect dipoles created by doping Ca and Mn are aligned with the crystal symmetry at thermodynamic equilibrium [60]. The restoring force created due to the random distribution of these defect dipoles accelerates the domain back-switching to its original state with zero net polarization. Also, from XRD the Ti and

Ba atoms displacement from the Ti-O plane and Ba-O plane of hexagonal phase in different directions also causes dipole alignment in multiple directions. From the XPS, we have already obtained that multiple oxidation states of $Mn^{3+}/Ti^{3+}$ concentrations are higher in BCTMO samples and hence the $O_v$. The amount of defect dipoles formed by positively charged ions with the $O_v$ will be more for the doped samples due to the larger proportions of $O_v$. These defects increase, and form clusters that resist the movement of domain walls, resulting in the pinching of the P-E loop. One must also note that sometimes aging of the samples too can be responsible for such pinched loops [61]. In these samples the aging-effect can be negated as the data has been taken from freshly prepared samples.

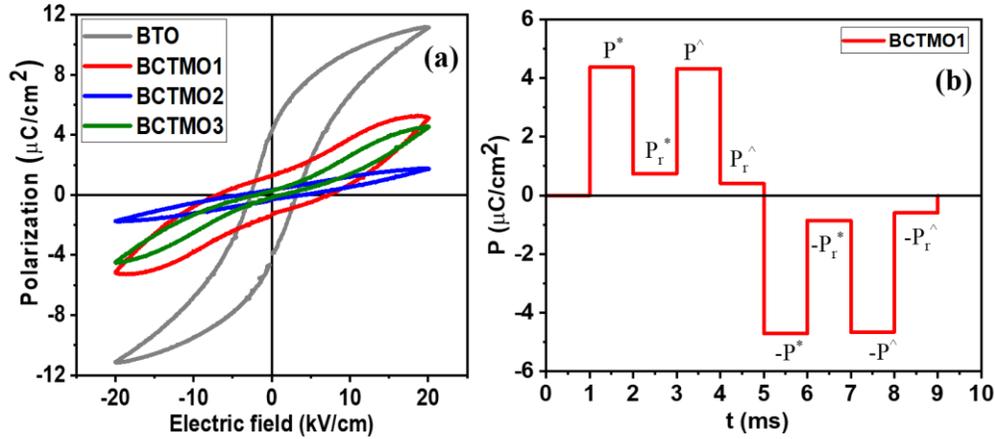

**Figure 14**: (a) P-E loop of all the samples (b) Measured polarization values at each pulse during PUND measurement.

Since the P-E loop of the doped samples are not saturated and showing a leaky behavior. To distinguish ferroelectric switching from artifacts for non-saturated P-E loop PUND (positive-up & negative-down) pulse measurement has been done. In this method a series of voltage pulses applied and measuring polarization by switching and non-switching voltages, that result allows for separation of the different components of the electrical response of a ferroelectric material. The details of the experimental part are provided in supplementary [SE2: PUND measurement].

The measurement has been done at voltage of 15kV with pulse width of 1 msec and the delay time of 1 msec. Switched and unswitched polarization is measured in both the positive and negative direction at both the pulse voltage and zero volts. Switched measurements are designate "P*" (P-Star) and unswitched measurements are called "P^" (P-Hat). Measurements at zero volts are distinguished by an appended "r". The eight measurements are, therefore, $\pm P^*$, $\pm P^*_r$, $\pm P^\wedge$ and

±P^r. The net switching polarization (dP) was evaluated using the following relation: (2P) = (±P* − (±P^). Figure 14 (a) shows the variation of saturated P-E loop for pure BTO to an unsaturated P-E loop for the doped samples. The results obtained from PUND for BCTMO1 samples is plotted in Figure 14b [PUND pulse plots for BCTMO2 and BCTMO3 are in Supplementary Figure S4 and values are tabulated in supplementary ST2]. The switching polarization obtained from the calculations are 0.0495 µC/cm$^2$ for BCTMO1, 0.0259 µC/cm$^2$ BCTMO2 and 0.0012 µC/cm$^2$ BCTMO3. Though the P-E loops of the doped samples are unsaturated and indicates leaky behavior, from the PUND measurement it has been proved that the samples are not completely leaky and retains proper ferroelectricity and contributes to the polarization exhibit by these materials.

**Magnetic properties**

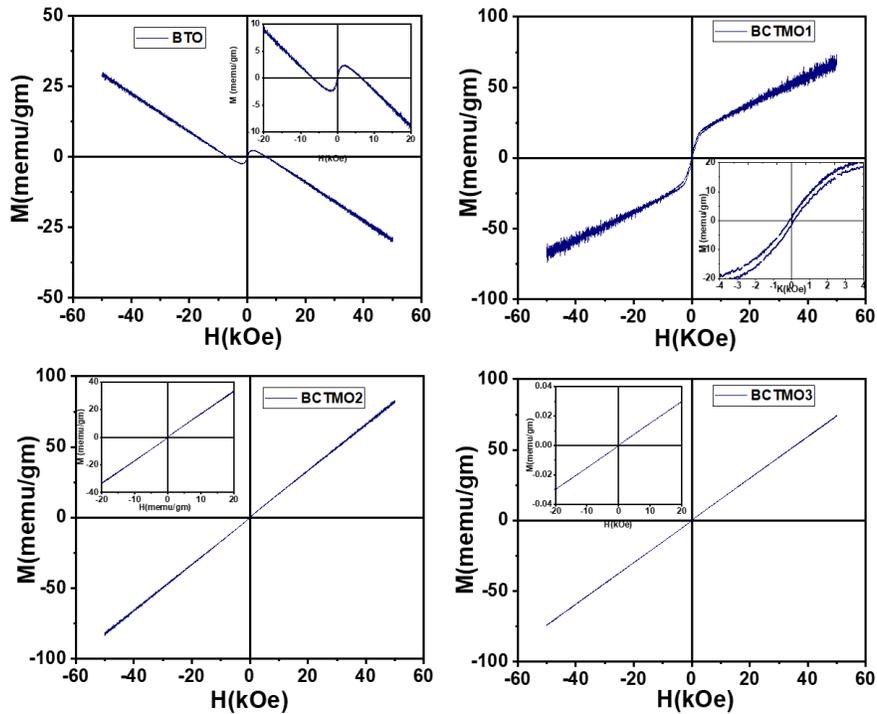

**Figure 15:** M-H loop of all the samples, inset shows the zoomed image at lower magnetic field

Incorporating Mn ions introduces magnetic properties in the materials, which can be realized from the M-H plots of the materials [Figure 15]. Pure BTO is diamagnetic, exhibiting a low negative magnetic moment with an applied magnetic field. This diamagnetic nature also confirms the Ti ions being in the $Ti^{4+}$ state having $d^0$ electrons. The 's' type nature of the diamagnetic loop indicates the feeble ferromagnetism arising due to the presence of $Ti^{3+}$. A weak ferromagnetic contribution in a predominantly paramagnetic background is observed for BCTMO1. The weak ferromagnetism is contributed due to $Mn^{3+}$-O-$Mn^{4+}$, $Ti^{3+}$-O-$Ti^{4+}$, $Mn^{3+}$-O-$Ti^{4+}$, $Ti^{3+}$-O-$Mn^{4+}$ double exchange (~180°) interactions [62]. Similarly, $Mn^{3+}$-O-$Mn^{3+}$, $Mn^{4+}$-O-$Mn^{4+}$, and $Ti^{3+}$-O-$Ti^{3+}$ superexchange (~90°) interactions also lead to weak ferromagnetism [6]. Among these double and super exchange mechanisms the Mn-O-Mn types are less probable due to the lesser proximity of the Mn ions. From the XRD analysis, the ~180° Ti/Mn-O-Ti/Mn bond angle was found to be the minimum for the BCTMO1 sample. This angle was ~167° for BTO, 164° for BCTMO1, 174° for BCTMO2 and 172° for BCTMO3. It is interesting to note that the ferromagnetic contribution is absent in BCTMO2 and BCTMO3 and the paramagnetic contribution becomes dominant in these samples. Hence, it appears that this angle may be critical for ferromagnetism. An increased hybridization of the 2p-3d orbitals in BCTMO1 may be also responsible for a lower angle thereby enhancing the ferromagnetic double exchange interaction. The comparison of the variation of M-H loop for all samples are plotted in figure 16 (a).

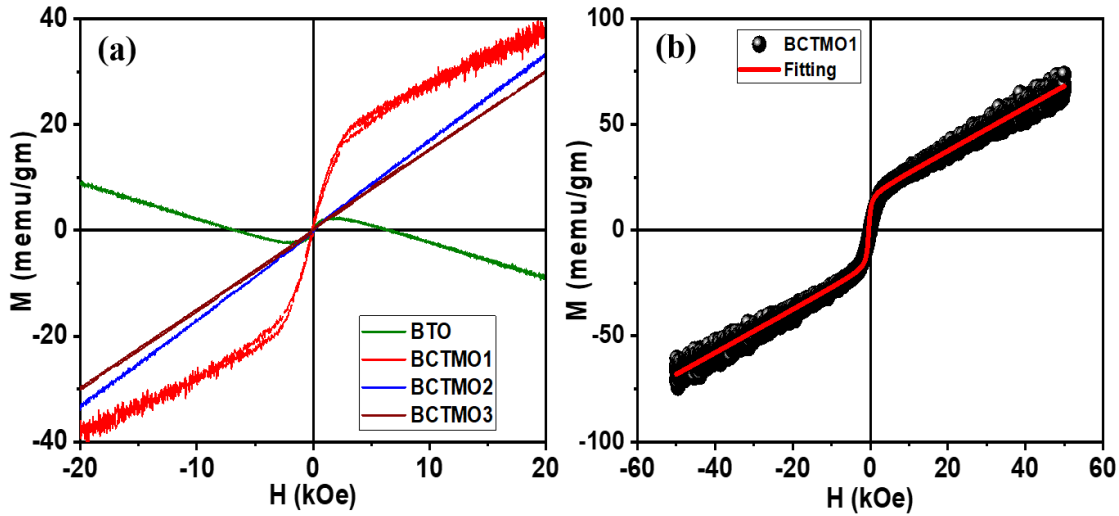

**Figure 16:** (a) M-H loop of all the samples, (b) Fitting of the M-H loop of BCTMO1 sample with ferromagnetic and paramagnetic component.

The M-H loop of the BCTMO1 sample can be separated into the sum of ferromagnetic and paramagnetic contribution by fitting the M-H curve with the following equation: [63] [Figure 16b]

$$M(H) = \left[2\frac{M_s}{\pi}\tan^{-1}\left\{\left(\frac{H \pm H_c}{H_c}\right)\tan\left(\frac{\pi M_r}{2M_s}\right)\right\}\right] + \chi H$$

The first term corresponds to the ferromagnetic contribution, and the second term corresponds to the paramagnetic contribution, where $M_s$ is the saturation magnetization, $M_r$ is remnant magnetization, and $H_c$ is the critical magnetic field. The fitted parameters are $M_s$ is $0.055 \pm 0.003$ emu/gm, $M_r$ is $0.01 \pm 0.001$ emu/gm, $H_c$ is $607 \pm 14$ Oe, and magnetic susceptibility, $\chi$ is $1.004 \times 10^{-6} \pm 2.3 \times 10^{-9}$ cm$^3$/gm. Hence, the modification of BTO to BCTMO1 can induce considerable ferromagnetism in the material along with the ferroelectricity.

**Magnetoelectric coupling in BCTMO1**

BCTMO1 shows both ferroelectric and ferromagnetic ordering indicating multiferroicity in the material [Figure 17 a, b]. However, existence of two ferroic orders does not ensure the coupling between the two. In this case a ferromagnetic and ferroelectric coupling, i.e. a magnetoelectric coupling is more desirable for practical applications than just mere presence of the two. The polarization, P and magnetization, M are related to the electric E and magnetic H fields by the following relations;

$$P_i = P_i^s + \varepsilon_0\varepsilon_{ij}E_j + \alpha_{ij}H_j + 1/2\beta_{ijk}H_jH_k + \gamma_{ijk}H_jE_k$$

$$M_i = M_i^s + \mu_0\mu_{ij}H_j + \alpha_{ij}E_j + 1/2\beta_{ijk}E_jE_k + \gamma_{ijk}E_jH_k$$

where spontaneous polarization and magnetization are represented by $P_i^s$ and $M_i^s$, respectively. $\varepsilon_{ij}$ is relative electrical permittivity and $\mu_{ij}$ is magnetic permeability of the substance and $\varepsilon_0$ and $\mu_0$ the permittivity and permeability of the free space. The magnitude of the linear ME effect in the material is determined by the tensor $\alpha_{ij}$ as $\alpha_{ij} = \delta P_i/\delta H_j$. Magnitude of the nonlinear ME effects are determined by the tensors $\beta_{ijk}$ and $\gamma_{ijk}$.

ME coupling can be investigate by the electric field generated in the material with the application of magnetic field. This electric field can be measured through the net charge accumulated on the electrodes, which can be estimated by shorting the two electrodes and measuring the net charge transfer from one electrode to the other through an external connector to which a charge-measuring unit was attached. To eliminate the influence of the electrical conductivity of the magnetic resistive component and electrode effect due to the direct magnetic field, ME effects are investigated dynamically by applying a combination of a steady (dc) and alternating (ac) magnetic fields to the samples. The dc field was varied from 0 to 1400 Oe in steps of 100 Oe, while an ac field of amplitude 40 Oe of frequency 10 Hz was applied. This enabled extracting information on the electric field, E generated due to an applied magnetic field, H. The magneto-electric coefficient, $\alpha_{ME} = \delta E/\delta H$ is measured using a Radiant Technologies Magneto-Electric Response Measurement unit using Vision software. The coefficient $\alpha_{ME}$ is related to $\alpha_{ij}$ as $\alpha_{ME} = \alpha_{ij}/\varepsilon_0\varepsilon_r$ [64].

The charge, Q generated by the application of the ac+dc magnetic field, H can be represented by: $Q = \beta H^2 + \gamma H - \xi$, where $\xi$, $\beta$ and $\gamma$ are constants obtained after fitting the P vs H curve. Note that $\delta Q/\delta H = 2\beta H + \gamma$. The charge Q can be converted by the software to the corresponding polarization, P, by using the equation, P=Q/A, where A = area of the pellet. Hence, one can obtain P v/s H data [Figure 17c]. The magnetoelectric voltage can be calculated using the relation: Q = CV, where, C = capacitance of the pellet = $\varepsilon A/t = \varepsilon_0\varepsilon_r A/t$, where, $\varepsilon$= dielectric constant of the material and t = the thickness of the pellet. The electric field, E can be expressed as E=V/t. Therefore, the ME coefficient, $\alpha_{ME}$ can be calculated using these parameters as:

$\alpha_{ME} = \delta E/\delta H = (1/t)\, \delta V/\delta H = (1/t)(1/C)\, \delta Q/\delta H = (1/t)(1/C)(2\beta H + \gamma) = (1/t)(t/\varepsilon_0\varepsilon_r A)(2\beta H + \gamma)$

Hence, $\alpha_{ME} = (1/\varepsilon_0\varepsilon_r A)(2\beta H + \gamma)$. To achieve $\alpha_{ME}$ the P-H data was fitted with regression analysis.

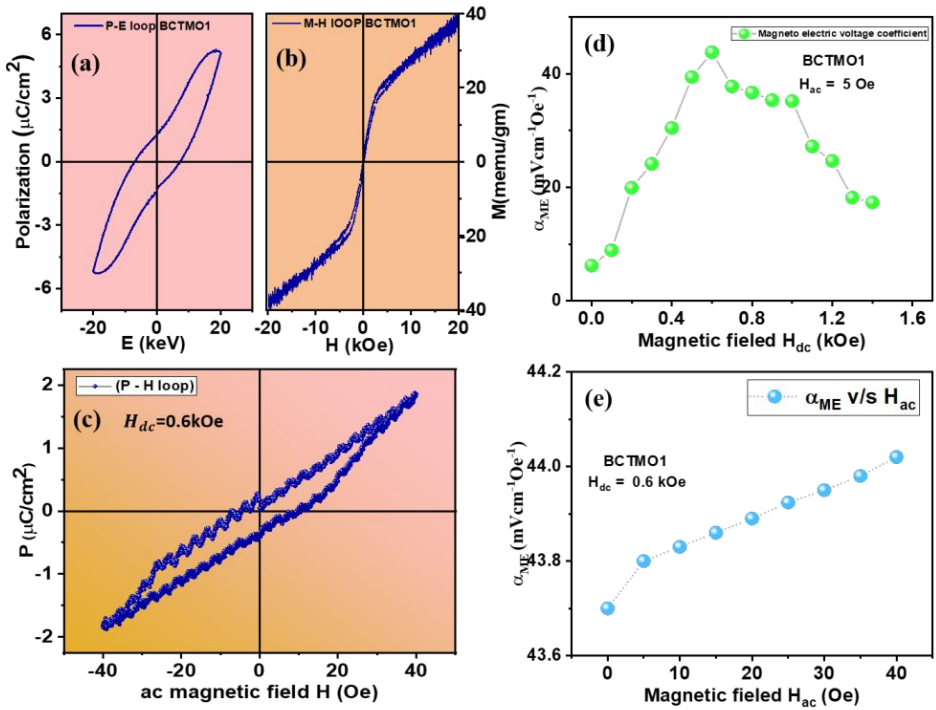

**Figure 17:** (a) P-E hysteresis of BCTMO1, (b) M-H hysteresis of BCTMO1, (c) P-H hysteresis of BCTMO1, (d) Variation of $\alpha_{ME}$ with dc magnetic field at constant ac magnetic field of 5 Oe, (e) Variation of $\alpha_{ME}$ with ac magnetic field at constant dc magnetic field of 0.6 kOe.

The magneto electric coupling coefficient, $\alpha_{ME}$ increases gradually with the increase in dc magnetic field with a constant ac magnetic field of 5 Oe. It attains a maximum value of 43.8 mVcm$^{-1}$Oe$^{-1}$ at 600 Oe dc field with 5 Oe ac field and there after gradually decreases to lower values of $\alpha_{ME}$ as the dc field increases [Figure 17d]. Magnetostriction of a material is the measurement of the amount of mechanical strain developed due to the application of a magnetic field resulting in the alignment of the magnetic domains. As the ME voltage arises due to magnetic-mechanical-electrical interactions, $\alpha_{ME}$ is directly proportional to the product of the piezomagnetic coefficient $q=d\lambda/dH$ (where $\lambda$ is the magnetostriction) and the piezoelectric coefficient d, $\alpha_{ME}$ is proportional to q×d. The dependence of $\alpha_{ME}$ with $H_{dc}$ is expected to follow the slope of $\lambda$ vs $H_{dc}$ of the magnetic material [65]. Hence the magnetoelectric coupling will be maximum where the magnetostriction is maximum. For a field greater than the maximum coupling field the polarization decreases. Such a scenario is observed in these materials, where a maximum coupling can be a resultant of maximum magnetostriction at about 500 to 1000 Oe, whereas for lesser and higher fields the coupling is weak and have lower values for $\alpha_{ME}$. Due to the application of the ac field (perturbation) there is a definite amount of movement in the ferromagnetic domains. The movement of these ferromagnetic domains are fluctuations in the spin orientation of the ferromagnetic components in the material. These fluctuations can affect the electronic clouds of the atoms in the vicinity of the ferromagnetic components. Such amendments to the electronic clouds are occasionally considerable to affect the structure such that the electric polarization changes accordingly, generating a modification of the strain in the structure. Note that, the $\alpha_{ME}$ increases linearly with the amplitude of $H_{ac}$ [Figure 17e] indicating the direct coupling of the ferroic orders. An increase of magnetoelectric voltage of 0.24 mVcm$^{-1}$Oe$^{-1}$ has been obtained for an increment of 40 Oe $H_{ac}$ at constant $H_{dc}$ of 600Oe [at 600 Oe $H_{dc}$, $\alpha_{ME}$ is 43.78 mVcm$^{-1}$Oe$^{-1}$ for 0 Oe $H_{ac}$ and 44.02 mVcm$^{-1}$Oe$^{-1}$ for 40 Oe $H_{ac}$]. Though this variation of the value of $\alpha_{ME}$ with ac magnetic field is very less, it proves the intrinsic coupling of the ferroelectric and ferromagnetic ordering within the material.

Hence, the incorporation of Ca and Mn in the BaTiO$_3$ is responsible for the structural transformation from a solely *P4mm* phase to a mixture of *P4mm* and *P6$_3$/mmc* phases with major *P4mm* phase (x=0.03,0.06) and with a major *P6$_3$/mmc* (x=0.09). For x=0.03, the magnetism induced in the sample is ferromagnetic in nature while with higher doping percentages the ferromagnetism weakens. The ferroelectricity in the materials also gets modified with incorporation of Ca and Mn. The loss of ferroelectricity along with the incorporation of magnetic properties, especially in the case of x=0.03 doping seems to be influencing each other in terms of a magnetoelectric coupling which is a consequence of magnetostriction in this material. The presence of magnetoelectric coupling in x=0.03 with ~12% *P6$_3$/mmc* phase, and the absence of a magnetic phase in higher doped samples with larger amount of *P6$_3$/mmc* phase is also indicative of the fact that an optimal amount of this phase may be instrumental in generating the magnetoelectric coupling.

**Conclusion**

Ba$_{(1-x)}$Ca$_{(x)}$Ti$_{(1-y)}$Mn$_{(y)}$O$_3$ (x= y= 0, 0.03, 0.06, 0.09), has been prepared through sol-gel synthesis. XRD and Raman spectroscopy reveals characteristics peaks corresponding to a dual phase of ferroelectric tetragonal *P4mm* and ferromagnetic hexagonal *P6$_3$/mmc* space groups of BaTiO$_3$. P-E loop shows the ferroelectric hysteresis for all samples and M-H loops indicates diamagnetism for x=0, ferromagnetic hysteresis for x=0.03 and paramagnetic behavior for other samples. The multiple oxidation states of Mn$^{3+}$, Mn$^{4+}$, Ti$^{3+}$, Ti$^{4+}$ and the amount of oxygen vacancies are the reasons for the ferromagnetism. A reduced O-Ti-O angle may be a consequence of an increased hybridization of the 2p-3d orbitals in x=0.03 sample as compared to other samples, which may be correlated to the enhanced ferromagnetic double exchange interaction. The existence of both ferroelectric and ferromagnetism in x=0.03 leads to multiferroicity and magnetoelectric coupling. A considerable magnetoelectric coupling coefficient of α$_{ME}$ ~44 mVcm$^{-1}$Oe$^{-1}$ was obtained for dc magnetic field of 600 Oe and an ac field of 40 Oe. The value of α$_{ME}$ increases linearly with the amplitude of H$_{ac}$ indicating the direct coupling of the ferroic orders. Hence, the Ba$_{(0.97)}$Ca$_{(0.03)}$Ti$_{(0.97)}$Mn$_{(0.03)}$O$_3$ is Magnetoelectrically coupled material. This ME coupling is absent for higher percentages of doping, revealing the adverse effect for higher percentages of Mn.

**Supplementary Material**

See the supplementary material for the explanation of sol gel synthesis [SE1] plots of XRD refinement data of all samples (S1), schematic made by VESTA software of the variation of Ti-Ti bond in octahedral dimer of P6$_3$/mmc space group (S2), XANES Pre-edge fitting using Athena software (S3), Experimental details of PUND measurement (SE2), PUND plots of BCTMO2 and BCTMO3 (S4), PUND results tabulated (ST1).

**Acknowledgement**

Authors MP, KSS and RS would like to thank the Ministry of Education, Government of India for the Prime Minister Research (PMRF) fellowship. The Authors would like to acknowledge the Department of Science and Technology (DST), Govt of India for providing the funds (DST/TDT/AMT/2017/200). The authors also acknowledge the Department of Science and Technology (DST), Govt. of India, New Delhi, India, for providing FIST instrumentation fund to the discipline of Physics, IIT Indore, to purchase a Raman Spectrometer (Grant Number SR/FST/PSI-225/2016). MP acknowledge for the Radiant technologies magneto electric precision tester installed at Materials Characterization Facility**,** Department of Physics & Materials Science Thapar Institute of Engineering and Technology Patiala - 147004, INDIA. The authors would like to acknowledge, the State University of New York (SUNY), Buffalo State University, and supported by the National Science Foundation, Launching Early-Career Academic Pathways in the Mathematical and Physical Sciences (LEAPS-MPS) program under Award No. DMR-2213412 for magnetic measurements.

**Data Availability**

The data supporting this article have been included as part of the Supplementary Information

# Supplementary material

# Room temperature Multiferroicity and Magnetoelectric coupling in

# (1-x) BaTiO$_3$-(x) CaMnO$_3$ solid solution

P. Maneesha[1], Koyal Suman Samantaray[1], Rakhi Saha[1], Tabinda Nabi[1], Rajashri Urkude[2], Biplab Ghosh[2], Arjun K Pathak[3], Indranil Bhaumik[4,5], Abdelkrim Mekki[6,7], Khalil Harrabi[6,8], Somaditya Sen[1]*

**Explanation SE1: Sol Gel synthesis**

Barium nitrate (purity 99.9%), Manganese nitrate (purity), Dihydroxy bis (ammonium lactate) Titanium (IV) 50% w/w aqua solution (purity 99.9 %), Calcium nitrate (purity = 99.9%), were used as precursors purchased from Alfa Aesar for sample preparation. The precursors were dissolved in DI water to form a clear solution. A solution of ethylene glycol (EG) as gelling agent and citric acid (CA) as fueling agent was prepared separately in the ratio 2:1 which was added to the precursor solution. The resultant solution (sol) was finally heated at 80°C. As the water evaporates, the sol gradually transforms to a gel. The self-burned gel were grounded and heated for denitrification at 450°C for 6 hr. further, decarbonized for 6 hours at 600 °C followed by calcination at 1200 °C for 3 hours. From the Phase formed powders required amount is mixed with Polyvinyl alcohol (PVA) and grounded well until becomes fine powders. This powder is pressed in to pellets using 10mm die set with 2-ton pressure for 3 minutes. These pellets are sintered at 1330 °C for 4 hours in a muffle furnace.

**Figure S1:**

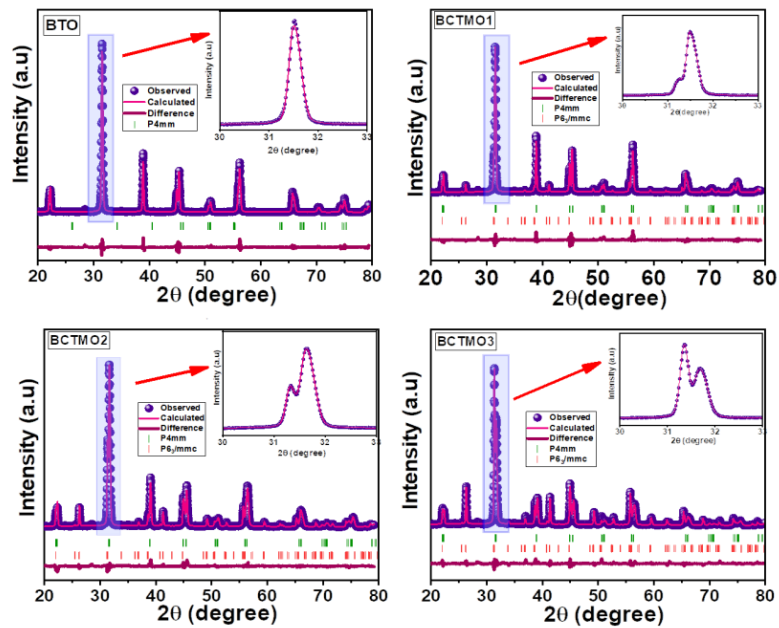

**Figure S1:** Refinement plots of XRD data of all samples. Inset shows the zoomed image of the emergence of hexagonal (104) peak near tetragonal (110) peak

**Figure S2:**

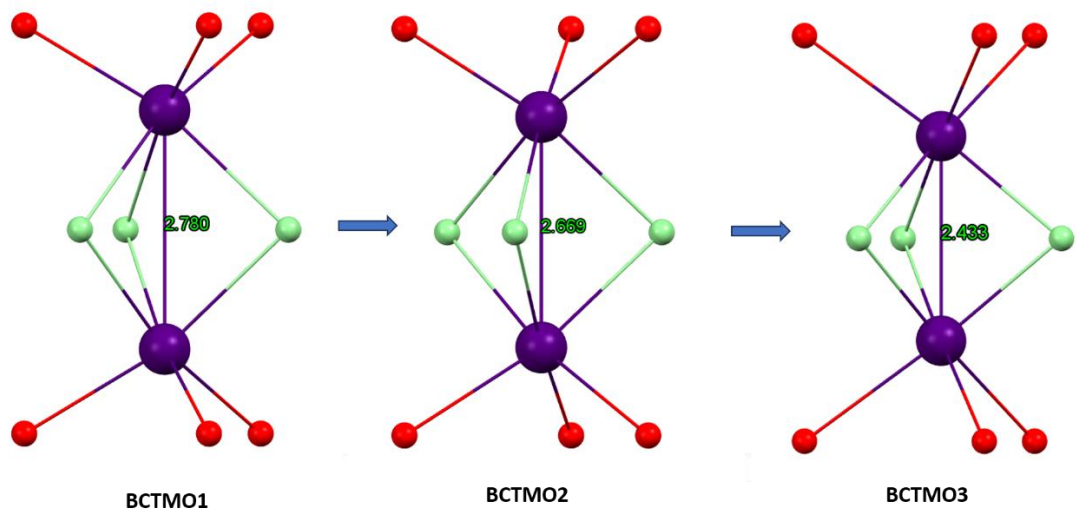

**Figure S2:** Variation of Ti-Ti bond in octahedral dimer of P6$_3$/mmc space group in doped samples.

**Figure S3:**

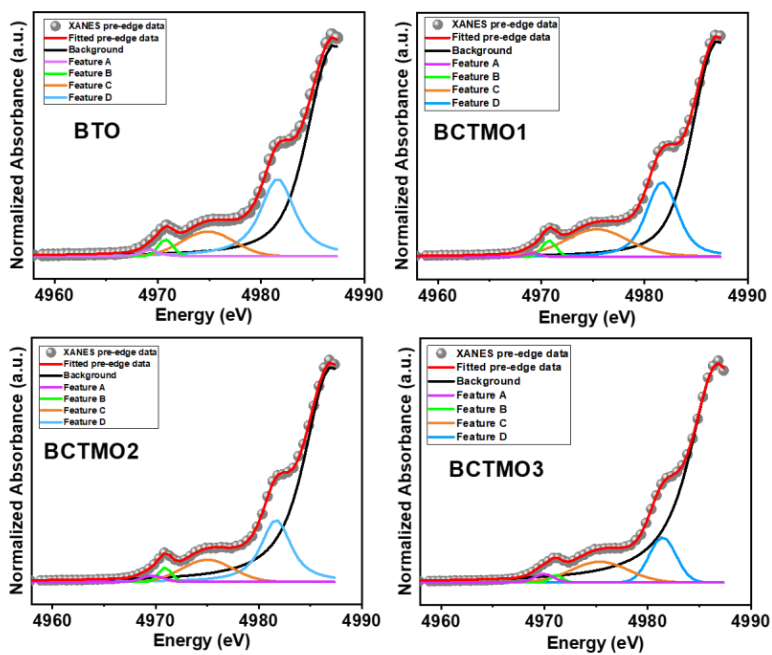

**Figure S3:** XANES Pre-edge fitting using Athena software

# Explanation SE2: PUND measurement

The PUND measurement is a standard ferroelectric test consisting of five pulses applied in sequence. The pulses are of the same (programmable) pulse width, with a fixed delay time between the pulses and are of the same magnitude ($|V_{Max}|$). The first pulse is in the negative $V_{Max}$ direction. It is not measured, but is used to preset the sample into the particular polarization ($\mu C/cm^2$) state. The next two pulses are in the positive $V_{Max}$ direction. The first switches the polarization and the second does not so that both switched and unswitched polarization are measured. At each pulse measurements are made with the pulse voltage applied after the pulse width and again after the voltage returns to zero and a delay of the pulse width (ms). The last two pulses are in the negative $V_{Max}$ direction with the first pulse switching the sample and the last pulse maintaining the switched state.

**Figure S4:**

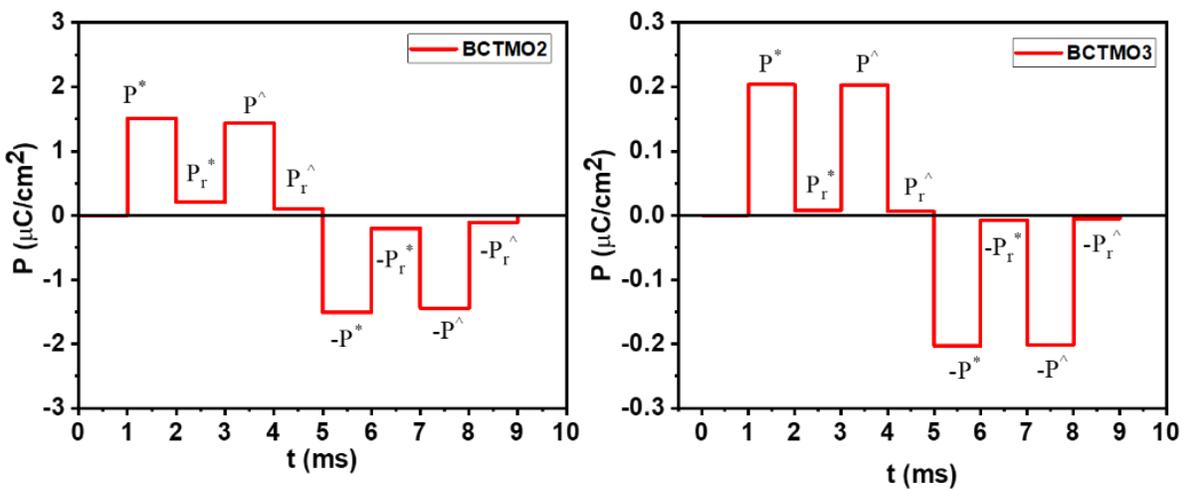

**Figure S4:** PUND measurement results of BCTMO2 and BCTMO3 samples

**Table ST1:**

| PUND measurement results of the BCTMO samples with unsaturated P-E loops ||||||
|---|---|---|---|---|---|
| Pulse | Applied voltage (kV) ($V_{max}$) | Measured values | BCTMO1 ($\mu C/cm^2$) | BCTMO2 ($\mu C/cm^2$) | BCTMO3 ($\mu C/cm^2$) |
| 1 | -15 | None | - | - | - |
| 2 | 15 | $P^*$ | 4.36908 | 1.51114 | 0.20365 |
|   | 0 | $P_r^*$ | 0.73448 | 0.20373 | 0.00794 |

| 3 | 15 | P^ | 4.32061 | 1.4952 | 0.20256 |
|---|---|---|---|---|---|
|   | 0 | $P_r^{\wedge}$ | 0.41005 | 0.10576 | 0.00619 |
| 4 | -15 | $P^*$ | -4.71032 | -1.50941 | -0.203 |
|   | 0 | $P_r^*$ | -0.8522 | -0.20289 | -0.00729 |
| 5 | 15 | P^ | -4.65976 | -1.47351 | -0.20169 |
|   | 0 | $P_r^{\wedge}$ | -0.59091 | -0.10943 | -0.00501 |
| Net Switching Polarization, dP | | | 0.04951 | 0.02592 | 0.0012 |